\documentclass[sigconf]{acmart}

\AtBeginDocument{%
  \providecommand\BibTeX{{%
    \normalfont B\kern-0.5em{\scshape i\kern-0.25em b}\kern-0.8em\TeX}}}

\copyrightyear{2025}
\acmYear{2025}
\setcopyright{cc}
\setcctype{by}
\acmConference[CHI '25]{CHI Conference on Human Factors in Computing Systems}{April 26-May 1, 2025}{Yokohama, Japan}
\acmBooktitle{CHI Conference on Human Factors in Computing Systems (CHI '25), April 26-May 1, 2025, Yokohama, Japan}\acmDOI{10.1145/3706598.3713288}
\acmISBN{979-8-4007-1394-1/25/04}

\usepackage{subcaption} 
\usepackage{url}
\usepackage{verbatim}
\makeatletter
\g@addto@macro{\UrlBreaks}{\UrlOrds}
\makeatother
\usepackage{multirow} 
\usepackage{color} 
\usepackage{enumitem} 

\usepackage{longtable}
\usepackage{todonotes}
\usepackage{xspace}

\newcommand{\m}{\textit{M=}}

\newcommand{\sd}{\textit{SD=}}

\usepackage{hyperref}

\MakeRobust{\ref}

\makeatletter
\newcommand{\labeltext}[2]{%
  \@bsphack
  \csname phantomsection\endcsname 
  \def\@currentlabel{#1}{\label{#2}}%
  \@esphack
}
\makeatother

\begin{document}

\title[Evaluating the Impact of Motion Fidelity on Optimized UI Design via BO in Automated UAM Simulations]{Fly Away: Evaluating the Impact of Motion Fidelity on Optimized User Interface Design via Bayesian Optimization in Automated Urban Air Mobility Simulations}
%


\author{Luca-Maxim Meinhardt}
\email{luca.meinhardt@uni-ulm.de}
\orcid{0000-0002-9524-4926}
\affiliation{%
  \institution{Institute of Media Informatics, Ulm University}
  \city{Ulm}
  \country{Germany}
}

\author{Clara Schramm}
\email{clara.schramm@uni-ulm.de}
\orcid{0009-0009-2297-5304}
\affiliation{%
  \institution{Institute of Media Informatics, Ulm University}
  \city{Ulm}
  \country{Germany}
}

\author{Pascal Jansen}
\email{pascal.jansen@uni-ulm.de}
\orcid{0000-0002-9335-5462}
\affiliation{%
  \institution{Institute of Media Informatics, Ulm University}
  \city{Ulm}
  \country{Germany}
}

\author{Mark Colley}
\email{m.colley@ucl.ac.uk}
\orcid{0000-0001-5207-5029}
\affiliation{%
  \institution{Institute of Media Informatics}
  \city{Ulm}
  \country{Germany}
}
\affiliation{%
  \institution{UCL Interaction Centre}
  \city{London}
  \country{United Kingdom}
}

\author{Enrico Rukzio}
\email{enrico.rukzio@uni-ulm.de}
\orcid{0000-0002-4213-2226}
\affiliation{%
  \institution{Institute of Media Informatics, Ulm University}
  \city{Ulm}
  \country{Germany}
}

\renewcommand{\shortauthors}{Meinhardt et al.}

\begin{abstract}
Automated Urban Air Mobility (UAM) can improve passenger transportation and reduce congestion, but its success depends on passenger trust. While initial research addresses passengers' information needs, questions remain about how to simulate air taxi flights and how these simulations impact users and interface requirements. \\
We conducted a between-subjects study (N=40), examining the influence of motion fidelity in Virtual-Reality-simulated air taxi flights on user effects and interface design. Our study compared simulations with and without motion cues using a 3-Degrees-of-Freedom motion chair. Optimizing the interface design across six objectives, such as trust and mental demand, we used multi-objective Bayesian optimization to determine the most effective design trade-offs.
Our results indicate that motion fidelity decreases users' trust, understanding, and acceptance, highlighting the need to consider motion fidelity in future UAM studies to approach realism. However, minimal evidence was found for differences or equality in the optimized interface designs, suggesting personalized interface designs.
\end{abstract}



\begin{teaserfigure}
\centering
 \includegraphics[width=0.99\textwidth]{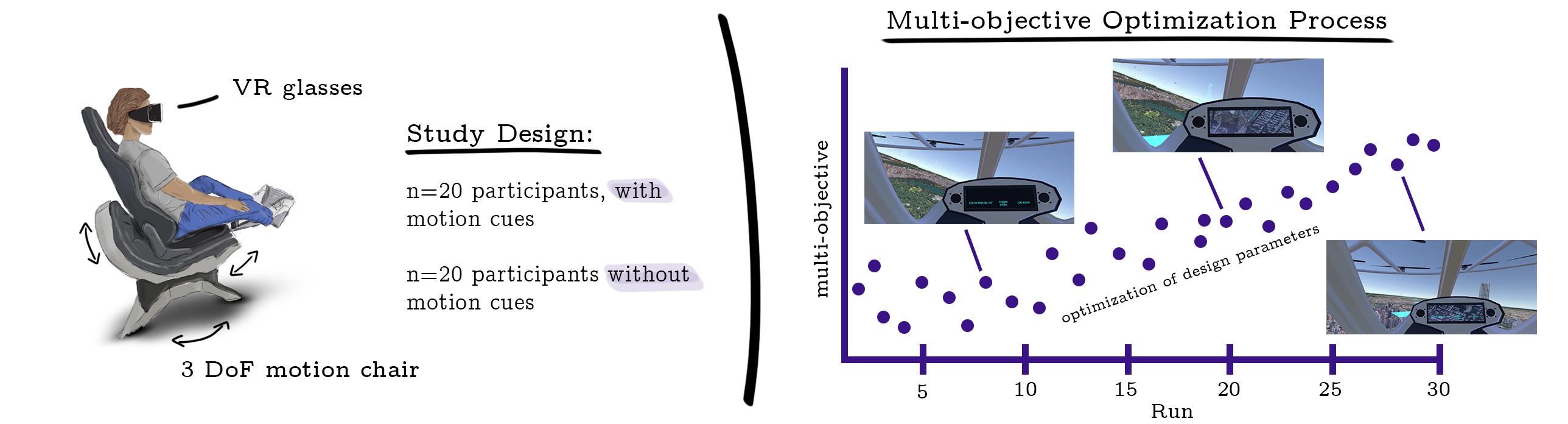}
  \caption{Overview of the user study setup and optimization process. (Left) N=40 participants were divided into two groups: n=20 experienced a simulated air taxi flight with motion cues using a 3-Degrees-of-Freedom motion chair and VR, while the other n=20 experienced the simulation in VR without motion cues. (Right) We applied Multi-Objective Bayesian Optimization to iteratively refine the air taxi’s user interface design over 30 optimization runs, focusing on six key objectives: trust, perceived safety, mental demand, understanding, acceptance, and aesthetics.}
  \Description{On the left side, there is an illustration of a person wearing VR glasses while seated in a 3 Degrees-of-Freedom motion chair, which moves to simulate the physical movements of an air taxi flight. Next to the image, the study design is described, noting that 20 participants experienced the simulation with motion cues, while another 20 participants experienced the simulation without motion cues. On the right side, the graph represents the multi-objective optimization process that was applied to optimize the user interface design. The x-axis shows the number of optimization runs (from 1 to 30), and the y-axis represents the aggregated multi-objective score. Dots on the graph represent the progression of the optimization over time, with improvement seen as the number of runs increases. Above the graph, there are three screenshots of the air taxi's UI. These images show different design iterations displayed on the simulated air taxi's UI, which provides flight information and trajectories.}
  \label{fig:teaser}
\end{teaserfigure}

\maketitle

\section{Introduction}
Automated Urban Air Mobility (UAM) is emerging as a potential solution to alleviate current congestion issues. For instance, in cities like London, Paris, and Brussels, people experience over 130 hours of congestion annually~\cite{Inrix.2021}. Even with the rise of hybrid working~\cite{HomeofficeUSA}, recent statistics show that traffic congestion continues to worsen~\cite{Inrix.2023}, highlighting the need for innovative transportation alternatives like UAM. 
Advancements in battery technology have enabled the development of electrically powered air taxis, which produce zero local emissions. For instance, Volocopter's electric air taxi, VoloCity, has a range of 35 km, sufficient for transporting passengers between city centers and airports in most of the world's megacities~\cite{VolocopterGmbH_WhitePaper}. These vehicles, known as electrical vertical take-off and landing (eVTOL) aircraft, are particularly beneficial due to their minimal urban space requirements. Due to these advances, the European Union Aviation Safety Agency (EASA) predicts that this new mode of transportation could become available within this decade~\cite{EuropeanUnionAviationSafetyAgency.2021}. 
In its early stages, UAM is expected to focus on small-scale operations, with air taxis carrying only a few passengers per trip~\cite{VolocopterGmbH_WhitePaper, EHangWhitePaper}. However, future visions extend beyond individual transportation, aiming to implement large-scale public transportation systems similar to buses~\cite{pak_can_2024}. Yet, in both visions, inexperienced, non-pilot passengers are expected to utilize these air taxis. Nonetheless, studies indicate that trust and perceived safety are critical factors influencing passengers' willingness to adopt to UAM~\cite{EuropeanUnionAviationSafetyAgency.2021, AlHaddad.2020, Edwards2020}. To enhance passenger trust, \citet{AlHaddad.2020} suggested implementing new safety standards, such as in-cabin surveillance cameras. Other solutions can be found in the field of Human-Computer Interaction, as Colley and Meinhardt et al.~\cite{colley2023comefly} demonstrated that simple trajectory visualizations on the windshield display (WSD) of air taxis significantly increase passengers' trust and perceived safety compared to no visualization or other visualizations such as a tunnel visualization or augmented landmarks. However, their study was only conducted as a low/mid-fidelity virtual reality (VR) study, although \citet{Edwards2020} recommended constructing a high-fidelity simulator to study passenger needs and the impact of rotor noise and vibration in the cabin. This raises the methodological question of how the fidelity of a simulator impacts the user studies' results when investigating passengers' information needs in the field of UAM. 

While most studies in the automotive domain have been conducted in VR~\cite{colley_effects_2021, fiore_redirected_2012}, recent research is already using higher-fidelity simulations like motion chairs~\cite{Meinhardt.2024} or real vehicles~\cite{flohr_prototyping_2023, kim_what_2023}. Comparing these different types of simulations, \citet{yeo_toward_2020} found significant differences in perceived motion fidelity and presence between VR-only setups, motion platforms, and real vehicles. Therefore, we suggest that motion fidelity may also affect users' information needs. However, to our knowledge, no studies have investigated the effects of motion fidelity on passengers in simulated UAM scenarios. This research is important, though, as the simulation's motion fidelity may influence the results' reliability, especially when addressing the complex issues of passengers' perceived safety and trust that arise for different visualizations of automated air taxis, as indicated by Colley and Meinhardt et al.~\cite{colley2023comefly}. While not focusing on motion fidelity, they only investigated a single aspect of visualization design for UAM--namely trajectories. However, a complete user interface (UI) involves multiple interconnected components. Optimizing the combination of these components becomes increasingly complex when trying to enhance multiple aspects, such as perceived safety and trust, which is especially critical in UAM contexts, where passengers often feel less familiar and more vulnerable compared to traditional ground transportation. Yet, in the field of Automated Vehicles (AVs), \citet{normark2015design} allowed passengers to manually customize the size, location, and color of icons on the dashboard, center stack, and Head-Up Display (HUD). By tailoring these UIs to individual preferences, the overall user experience can be enhanced~\cite{sun_exploring_2020}, fostering increased perceived safety and trust in future automated UAM.

Instead of relying on manual adjustments to various design elements within the visualizations, Multi-Objective Bayesian Optimization (MOBO) offers a more efficient solution. MOBO iteratively identifies the optimal design parameters by incorporating user feedback at each iteration, progressively refining the design to meet the users' needs. This approach has already been successfully applied in various domains to solve design optimization problems~\cite{chan2022bo, liao2023interaction, chandramouli2023mobopersonalize, koyama2022boassistant, kadner2021adaptifont}. MOBO predicts which design changes will most effectively achieve the desired objectives, such as increasing passengers' trust and perceived safety. It manages multiple objectives by identifying the best trade-offs among them, known as the Pareto front, ensuring the most effective compromises for UI design~\cite{marler_survey_2004}.
In light of these considerations, this work is guided by the following two research questions (RQs):

\begin{quote}
\smallskip

\noindent\fcolorbox{black}{black!10}{\textbf{RQ1}} \textit{What are the characteristics of an optimized UI design for automated air taxis that enhance passengers' trust, perceived safety, mental demand, understanding, acceptance, and aesthetics?}

\smallskip
\smallskip

\noindent\fcolorbox{black}{black!10}{\textbf{RQ2}} \textit{How does the motion fidelity of the simulation affect the design parameters and user's effects for a UI in automated UAM?}

\smallskip
\end{quote}

To address these RQs, we conducted a between-subject user study with N=40 participants. The participants were divided into two groups, with n=20 participants, each experiencing a simulation of an automated air taxi flight created in Unity. The first group experienced the simulation using a 3-degrees-of-freedom (DoF) motion chair combined with VR, while the second group experienced the simulation using VR only, without the motion chair.
Additionally, we applied MOBO to the air taxi's UI design. This optimization focused on 12 design parameters, such as the length and transparency of the ego-trajectory on the WSD, a map visualization on the internal display, and the visualization of boundary boxes around other air taxis. We optimized these parameters based on six objectives: trust in automation, understanding, mental demand, perceived safety, acceptance, and aesthetics.

Our findings show that while motion fidelity had a minimal effect on most UI design parameters, we found strong to extreme evidence that it reduces users' trust, understanding, and acceptance, highlighting its importance in future UAM studies. Nonetheless, the lack of differences in immersion between the groups may derive from participants' limited experience with real air taxi flights. Additionally, the relatively high variance in design preferences on the Pareto front suggests that a one-size-fits-all approach may not be suitable for air taxi interfaces. Although objectives improved during optimization, personalized interfaces tailored to individual preferences could enhance user experiences. Eventually, we provide practical guidelines for future UAM research on automated air taxi interfaces.

\smallskip
\smallskip
\smallskip

\noindent\fcolorbox{purple}{purple!20}{\textbf{Contribution Statement}~\cite{Wobbrock.2016}}
\begin{itemize}[noitemsep]
    \item \textbf{Artifact or System} We developed a VR, Unity-based simulation of an automated air taxi flight, designed to iteratively optimize the UI through MOBO, allowing the identification of optimal design parameters based on user feedback across six objectives.
    
    \item \textbf{Empirical study that tells us about how people use a system.} We conducted a between-subjects study (N=40) to investigate the impact of motion fidelity on user experience and UI design for automated air taxis. Our findings show that there is evidence that motion fidelity decreases trust, understanding, and acceptance while having minimal effect on optimized UI design, suggesting the need for tailored user interfaces in future UAM applications.

\end{itemize}





\section{Related Work}

This section will provide an overview of prior research on users' information needs within the context of UAM and explore potential UIs designed to visualize this information. Furthermore, it introduces Bayesian Optimization (BO) as an algorithm for optimizing these UIs.

\subsection{Human-Computer Interaction for Urban Air Mobility}
Approach UAM with an HCI perspective, \citet{Kim.2021}, conducted a workshop focused on User Experience in UAM with HCI experts. The insights from this workshop were further analyzed by \citet{Lim.2022}, revealing a transition in focus from initial concerns of \textit{safety} and \textit{acceptance} in the first phase of UAM to later emphasizing \textit{comfort} as potential passengers' major concerns. However, they did not propose explicit solutions for these concerns.
\citet{Edwards2020} conducted a similar workshop with aviation professionals, identifying six primary categories of passenger concerns: safety, noise and vibration, passenger well-being, and environmental concerns. To address these concerns, they suggested creating a high-fidelity simulator to understand passenger needs better and study the effects of rotor noise and vibration inside the cabin. 
Further advancing the field, Meinhardt and Colley et al.~\cite{meinhardt2023up} held a workshop with six professional helicopter pilots to evaluate automation and visualization possibilities for future UAM piloting. Their focus included visualizing air traffic, avoiding obstacles, and map visualization to enhance situational awareness. Their findings align with \citet{janetzko_what_2024}, who investigated the information needs of emergency flight pilots, noting a desire to prioritize critical information, such as collision avoidance, during these flights. 
Diving into the visualization of these information needs, initial research was conducted by Colley and Meinhardt et al.~\cite{colley2023comefly}. In a VR user study, they found that a chevron path line visualization was the most effective visualization for significantly increasing trust and perceived safety compared to other visualizations. Particularly in the presence of other air traffic, the path line increased trust and provided a more predictable view of the air taxi's future trajectory. Based on their work, \citet{valente_towards_2024} investigated the path line visualization during different flight phases (take-off, cruise, landing) and visibility conditions (daylight, night, foggy). They found that their participants' understanding decreased during take-off and landing and under poor visibility conditions when using visualizations other than the path line, suggesting that this visualization performs best in most tested conditions.

Drawing from the insights of visualization of passengers' information needs~\cite{colley2023comefly, valente_towards_2024, meinhardt_wind_2024}, we see that \textit{how} flight-relevant information is displayed significantly impacts passengers' trust and perceived safety. However, these studies have primarily focused on a single visualization component--namely trajectories, whereas a complete UI involves multiple interconnected components. Optimizing the combination of these components, however, becomes increasingly complex. To address this challenge, this work employs MOBO to explore and optimize multiple parameters in the design of air taxi UIs. Therefore, the next section introduces MOBO for interface designs to relate to \fcolorbox{black}{black!10}{\textbf{RQ1}} of this research.

\subsection{Multi-Objective Bayesian Optimization for User Interface Designs}
Designing an effective UI for automated UAM involves selecting the best combination of design parameters—such as position, transparency, and size—to achieve key objectives like enhancing trust and perceived safety. Given the complexity of modern UIs, where numerous design parameters must be optimized, testing every possible combination is infeasible.

BO offers an efficient solution by modeling the relationship between design parameters and objective functions. BO iteratively refines designs by balancing seeding (trying new designs) and exploitation (focusing on promising ones), which makes it particularly effective in optimizing UI designs with limited data and unknown outcomes~\cite{brochu2010tutorial}. 
For subjective design objectives, such as trust and perceived safety, it is necessary to integrate the user into the BO process. In such a Human-in-the-Loop (HITL) process (see \citet{chan2022bo}), the BO adjusts the design parameters in each optimization iteration based on the user's subjective evaluation of the last design.
HITL BO has been applied successfully in various UI design challenges, such as defining animation parameters based on user feedback~\cite{brochu2010bayesian}, optimizing font settings for faster reading~\cite{kadner2021adaptifont}, and improving UI interactions to reduce task completion times~\cite{dudley2019crowdsourcing}.
For UI design for automated air taxis, multiple objectives like trust, perceived safety, and mental demand must be balanced~\cite{colley2023comefly, valente_towards_2024, meinhardt_wind_2024}. Hence, MOBO (e.g., see \citet{johns_towards_2023}) can be used to find the optimal design parameters that balance all objectives. In particular, MOBO generates a range of optimal designs, known as the Pareto front, where no design can improve one objective without compromising another~\cite{marler_survey_2004}. This approach has been applied in contexts like touchscreen keyboards~\cite{dunlop2012multidimensional} and haptic interfaces~\cite{hayward1994design}.

In the context of UAM, the UI design challenge is further complicated by the need to account for the impact of motion fidelity on user perceptions. Unlike automated driving, where motion cues occur primarily in two dimensions, UAM involves three-dimensional movement, including vertical maneuvers like takeoffs and landings. Research has shown that passengers' understanding is lower during vertical flights compared to cruising~\cite{valente_towards_2024}. However, this research was based on video simulations that did not incorporate actual motion, raising questions about how motion fidelity might affect the study outcomes in the UAM domain. To address these concerns, we explored how different levels of motion fidelity (motion cue and no motion cues) influence UI design in UAM, as outlined in \fcolorbox{black}{black!10}{\textbf{RQ2}}. The following section will give an overview of motion fidelity simulators in the field of AVs and UAM.

\subsection{Motion Fidelity Simulators}\label{subsec:motionfidality}
Simulators used in AV studies vary widely in terms of motion fidelity, which refers to how accurately they replicate the physical movements of a vehicle. This fidelity can range from fixed-base setups with minimal or no motion cues to advanced moving-base simulators with multiple degrees of freedom (DoF)~\cite{mohajer_VehicleMotionSimulators_2015, bruck_ReviewDrivingSimulation_2021, hock_introducing_2022}. 

Low-cost, fixed-base simulators employ simple rotational or linear motion to simulate basic vehicle dynamics~\cite{danieau_HapSeatProducingMotion_2012a, bouyer_InducingSelfmotionSensations_2017, hock_ElaboratingFeedbackStrategies_2016}. In contrast, high-end moving-base simulators offer more sophisticated motion representation, such as 6-DoF platforms that can simulate complex maneuvers including pitch, roll, and yaw~\cite{chen_NADSUNIVERSITYIOWA_2001, colleySwiVRCarSeatExploringVehicle}.
Recent works focused on achieving effective motion fidelity with more accessible technologies in terms of costs. For example, the \textit{SwiVR-Car-Seat} developed by \citet{colleySwiVRCarSeatExploringVehicle} uses a 1-DoF rotational seat to simulate longitudinal and lateral vehicle dynamics. This approach is based on the finding by \citet{rietzler_VRSpinningExploringDesign_2018} that angular impulses alone can effectively represent vehicle acceleration or deceleration. Building on this concept, Hock and Colley et al.~\cite{hock_introducing_2022} simulated AV motion using a remote-controlled wheelchair, leveraging its acceleration and rotation capabilities.
For comparison, \citet{yeo_toward_2020} developed six simulators with varying levels of motion fidelity, ranging from monitor setups without motion cues to mixed reality environments combined with a real vehicle. Their study revealed that this actual vehicle, combined with a mixed reality headset, had the significantly highest perceived motion fidelity and perceived presence. However, this research primarily focused on the fidelity aspect without exploring how different levels of motion fidelity might influence the effects of UIs on users in these environments.

In the context of UAM, motion simulators for air taxi flights have also been developed, utilizing motion chairs~\cite{bergman2018mixed} or CAVE virtual environments~\cite{marayong_urban_2020}. However, these studies primarily concentrate on constructing the simulators rather than examining how different levels of motion fidelity might affect user preferences for UI design during flights. To address this gap and explore the optimization of UIs in automated UAM, we conducted a user study detailed in the following section.

\section{User Study}
To evaluate the impact of different UI optimizations in VR environments, both with and without motion cues, we conducted a between-subject design study with N=40 participants. The participants were divided into two groups: one group experienced the simulation without motion cues (n=20), while the other group used a 3-DoF motion chair (n=20).

\subsection{Apparatus}
We developed a simulation using Unity~\cite{unitygameengine} version 2022.3.19f1 to present the ego perspective of a passenger inside the Volocopter 2X flying over New York City. We utilized the Google Map Tiles API for Unity to render the environment. In the simulation, the Volocopter followed a predetermined route that included various maneuvers, such as turns in both directions and altitude changes. We incorporated other air taxis into our simulations, as their presence influences passengers' trust and perceived safety~\cite{colley2023comefly, meinhardt_wind_2024}. Predictions for New York suggest that approximately 500 air taxis might operate at any given time~\cite{Haan.2021, Pukhova.2021, Rajendran.2020, Rajendran.2019}. Consequently, our simulation featured 500 air taxis operating on randomly timed predefined trajectories.
Forty participants experienced the simulation, split into two groups. One group of n=20 participants used the Vive Pro headset without a motion chair, while the other group of n=20 participants used the Vive Pro together with the YAW VR motion chair~\cite{yawvr}, which features 3 DoF. The chair's motion was synchronized with the movement of the virtual air taxi, providing a 1:1 translation of the air taxi's rotation (yaw, pitch, and roll).
In line with the work of Colley and Meinhardt et al.~\cite{colley2023comefly}, the air taxi was equipped with a WSD capable of visualizing the trajectory of both the ego air taxi and the other air taxis. Additionally, the WSD highlighted other air taxis using boundary boxes, a technique inspired by prior research in the automotive domain where recognized vehicles are similarly highlighted~\cite{Woide.2022, colley2021effects, colley2022effects}. The air taxi also featured a display as the dashboard. It showed a map of the environment beneath the ego air taxi, including its trajectory and additional information such as current speed and altitude (see \autoref{fig:optimized_parameters}).

\subsection{Multi-Objective Bayesian Optimization}
The Unity simulation was connected to a Bayesian multi-objective optimizer from the Python package BoTorch~\cite{balandat2020botorch} (version 0.11.1). This optimizer was used to iteratively adjust 12 design parameters of the UI, such as the length of the ego trajectory and its transparency (the following section will explain the design parameters in detail). We employed the \texttt{qEHVI} acquisition function, representing the expected hypervolume increase. To ensure that after each run, only a single batch is selected for evaluation, we set q=1, in line with \citet{chan2022bo}.

The optimization process began with a five-run sampling phase using Sobol sampling \cite{sobol1967}. This method systematically divides the design space into evenly distributed regions and selects representative design parameter configurations for each. By collecting user ratings on these samples, the optimizer explores how the individual user perceives broader regions of the design space. This identifies the initial promising regions for further exploration. All users received the same five samples during the sampling phase to prevent bias arising from different starting points, which represents the default BoTorch behavior. We deemed five samples appropriate based on internal tests and prior studies on air taxi UIs, which identified significant user preference differences with a few distinct designs~\cite{valente_towards_2024,meinhardt2023up}. This suggests that Pareto-optimal designs can likely be discovered in a comparable number of designs' origin regions, reducing the need for extensive seeding in the first phase.
In the following 25-run optimization phase, the acquisition function dynamically balanced ''exploitation'' (refining known \textit{good} configurations) and ''seeding'' (searching for new regions of the design space). In line with previous work using MOBO \cite{chan2022bo,shahriari2015taking}, the acquisition function governs this balance, as users unfamiliar with an optimization process might not know which balance to choose manually (This is in contrast to \cite{zhou_interactive_2021}, which explores user-driven balancing). 
We used 1024 restart candidates and 512 Monte Carlo samples to approximate the acquisition function during the optimization process. The design parameters of the UI inside the air taxi were updated throughout this process. 
By having more optimization runs compared to sampling runs, we focused on fine-tuning the design parameters based on individual user feedback. This aimed to converge more efficiently to personalized designs.

\subsection{Design Parameters}
\begin{figure*}[ht!]
    \centering
    \small
    \includegraphics[width=0.98\linewidth]{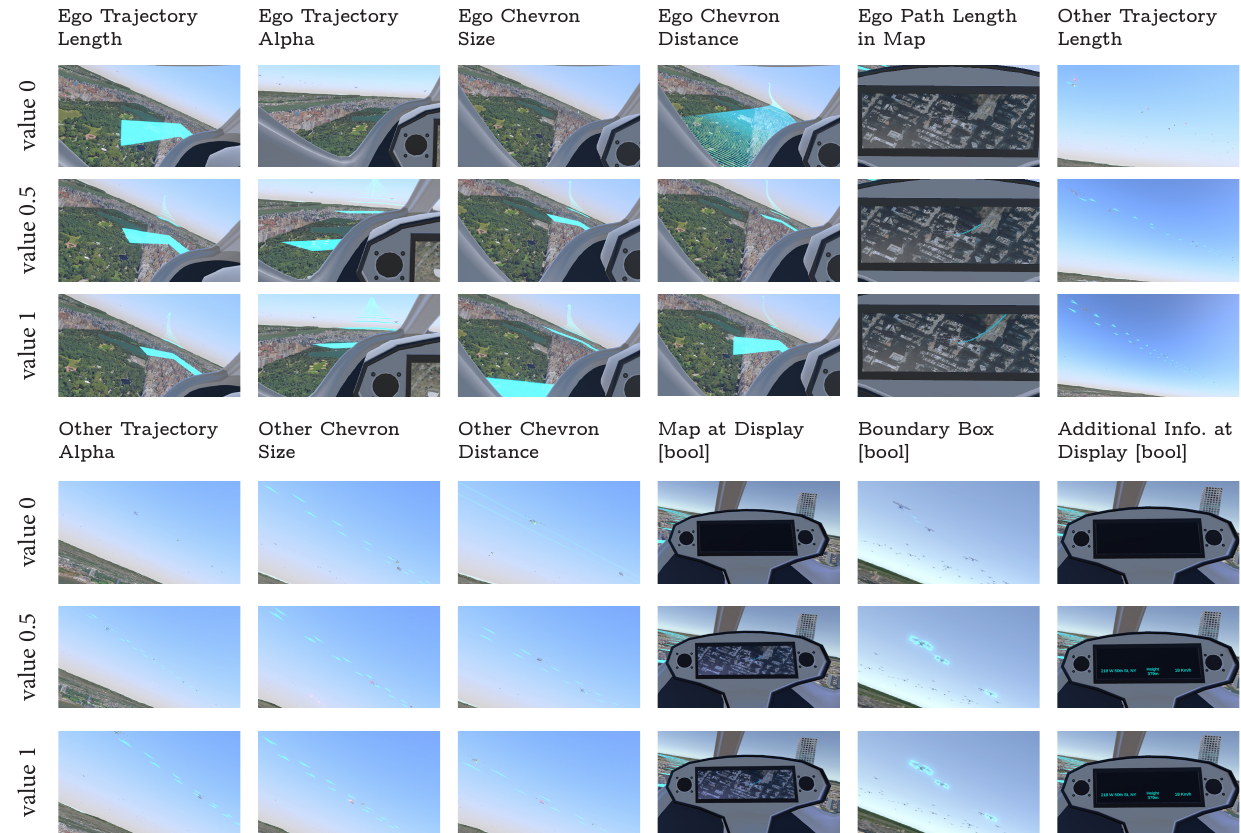}
    \caption{Design parameters to be optimized. The figure shows the values of 0, 0.5 and 1 for each design parameter}\label{fig:design_parameters_show}
    \Description{The figure depicts the 12 design parameters to be optimized for the air taxi interface. A horizontal group of pictures represents each parameter's values from 0, 0.5, and 1. The parameters include Ego Trajectory Length, Ego Trajectory Alpha, Ego Chevron Size, Ego Chevron Distance, Ego Path Length in Map, Other Trajectory Length, Other Trajectory Alpha, Other Chevron Size, Other Chevron Distance, Map at Display (bool), Boundary Box (bool), and Additional Info at Display (bool). The values represent different levels for each parameter.}
\end{figure*}

The 12 design parameters to be optimized were derived from related work in the field of UAM~\cite{valente_towards_2024, colley2023comefly, motnikar_unravelling_2023}. All parameters were mapped to a range between 0 and 1. While most parameters were on a continuous scale, three were binary, representing whether a UI element was visible. To handle these binary parameters, we set a threshold: elements with a value of less than 0.5 were considered invisible, while those with a value of 0.5 or more were considered visible. This approach was chosen because MOBO is typically more efficient when working with continuous parameters~\cite{shahriari2015taking}. 
The color was deliberately excluded as a design parameter, as turquoise for all augmented objects in the WSD follows an established design principle~\cite{werner_new_2018}, ensuring a neutral appearance. Further, the visualization range in the WSD is capped at 1 km; beyond this distance, path visualizations and boundary boxes are not displayed, as they would no longer be perceivable.
In \autoref{fig:design_parameters_show}, all parameters to be optimized are depicted, including their range between 0 to 1.

\textbf{Design Parameters:}
\begin{itemize}[leftmargin=*, itemsep=0pt]
\item \textbf{Ego Trajectory Length:} \\The length of the ego trajectory path in the WSD, scaled between 0m and 1km (maximum clipping distance).
\item \textbf{Ego Trajectory Alpha:} \\The transparency of the ego trajectory path in the WSD.
\item \textbf{Ego Chevron Size:} \\The size of the chevrons along the ego trajectory path, scaled between 0m and 12.5m.
\item \textbf{Ego Chevron Distance:} \\The distance between individual chevrons, scaled between 0m and 42m.
\item \textbf{Ego Path Length in Map:} \\The length of the ego path displayed on the map represents a range from 0m to 260m (260m being fully visible on the entire display).
\item \textbf{Other Trajectory Length:} \\The length of the trajectory paths for other air taxis, scaled between 0m and 205m, corresponding to the distance between these air taxis in the simulation.
\item \textbf{Other Trajectory Alpha:} \\The transparency of the trajectory paths for other air taxis in the WSD.
\item \textbf{Other Chevron Size:} \\The size of chevrons along the trajectory paths of other air taxis, scaled between 0m and 12.5m.
\item \textbf{Other Chevron Distance:} \\The distance between individual chevrons on the trajectory paths of other air taxis, scaled between 0m and 42m.
\item \textbf{Map Display:} \\A binary parameter indicating whether a map is shown at the display ($<0.5$: no map, $\geq0.5$ map is displayed).
\item \textbf{Boundary Box:} \\A binary parameter indicating whether the boundary boxes are displayed on the WSD ($<0.5$ no boundary box, $\geq0.5$: boundary boxes displayed
\item \textbf{Additional Information on Display:} \\A binary parameter indicating whether additional information (altitude and speed) are shown at the display ($<0.5$ no boundary box, $\geq0.5$: boundary boxes displayed)
\end{itemize}

\subsection{Objective Function}\label{sec:objective_functions}
An objective function \( f \) maps a visualization design \( x \) to a subjective metric that the optimizer seeks to maximize or minimize. In our study, we focused on six subjective metrics informed by previous research~\cite{colley2022scene,colley2023comefly}: \textit{perceived safety}, \textit{trust in automation}, \textit{understanding}, \textit{mental demand}, \textit{acceptance}, and \textit{aesthetics}. Among these, \textit{mental demand} was the only metric we aimed to minimize, while the others were to be maximized. To evaluate these metrics throughout the optimization process, we employed established questionnaires, as presented in the following:

\textbf{Mental demand} was assessed using the mental workload subscale of the raw NASA-TLX~\cite{hart1988development} on a 20-point Likert scale. \textbf{Understanding} was measured via the \textit{Predictability/Understandability} subscale from the \textit{Trust in Automation} questionnaire by \citet{korber2018theoretical}. \textbf{Trust} was evaluated using the \textit{Trust} subscale from the same questionnaire, both using a 5-point Likert scale. \textbf{Perceived safety} was rated by participants using four 7-point semantic differentials ranging from -3 (anxious/agitated/unsafe/timid) to +3 (relaxed/calm/safe/confident), where a higher score indicates greater perceived safety~\cite{faas2020longitudinal}.
Additionally, we included three single items to assess specific aspects of the visualization design. \textbf{Acceptance} was assessed using two items inspired by the van der Laan acceptance scale~\cite{van1997simple}: “I find the visualizations of the automated vehicle \textbf{useful}” and “I find the visualizations of the automated vehicle \textbf{satisfying}.” These were averaged into a single acceptance objective. \textbf{Aesthetics} was adapted from \citet{colley2023comefly} and measured visual appeal with the statement, “I found the visualizations visually appealing,” rated on a 7-point Likert scale.

To ensure consistency across the questionnaires, which employed different Likert scales, we normalized all six metrics to a range of $[-1, 1]$. Notably, for the mental demand metric, we inverted its orientation such that higher normalized values indicate lower perceived mental demand.

\subsection{Procedure}
The procedure was similar for both conditions (VR with and without a motion chair). Each session began with a brief introduction, followed by participants agreeing to a consent form. Participants then experienced 30 seconds of a simulated air taxi flight. Afterward, they rated their experience using the subjective metrics detailed in \autoref{sec:objective_functions}. The Bayesian Optimizer used these ratings to update the design parameters and present the next iteration of the UI design. Each participant went through 30 optimization runs in total. However, if a participant gave the highest possible ratings for all objectives (or the lowest for mental demand) for a specific UI design, the study ended early, as this was considered the optimal design for that participant. After all runs, a demographic questionnaire was administered, asking questions about the participant's age, gender, and immersion using the subscale of the Technology Usage Inventory (TUI)~\cite{kothgassner2013technology}.
For their 1.5h effort, the participants were compensated with 15\anon{Euros}.

\subsection{Participants}

As we conducted a between-subject study, the participants' demographics are reported separately for each group.

The group with motion cues had an average age of 25.60 years (\sd{3.26}). This group consisted of 10 males, 10 females, and no non-binary participants. Among them, four participants were employed, one was self-employed, and the remaining were college students. The group without motion cues had an average age of 24.95 years (\sd{2.99}). This group included 11 males, 9 females, and no non-binary participants. All participants in this group were college students.

\subsection{Results}

\subsubsection{Analysis}
The goal of MOBO is to identify the Pareto front, which consists of all Pareto optimal points in the design space. Each point on this front (e.g., a combination of design parameters in a UI) represents a design that cannot be improved in one objective without compromising another. This set of optimal designs captures the most efficient trade-offs between conflicting design objectives~\cite{marler_survey_2004}. Hence, we determined the Pareto optimal values using the R package \textit{EMOA: Evolutionary Multiobjective Optimization Algorithms}~\cite{emoa} for each participant. For the subsequent analysis, we exclusively used these Pareto values, filtering the combinations of design parameters and questionnaire ratings to include only those located on the Pareto front. This approach ensures we only considered the most efficient and balanced designs in our subsequent analysis. While the group experiencing motion during the study yielded n=48 Pareto designs, the group only experiencing the study in VR without motion yielded n=42 Pareto designs.

\begin{figure}[ht!]
    \centering
    \small
    \includegraphics[width=0.99\linewidth]{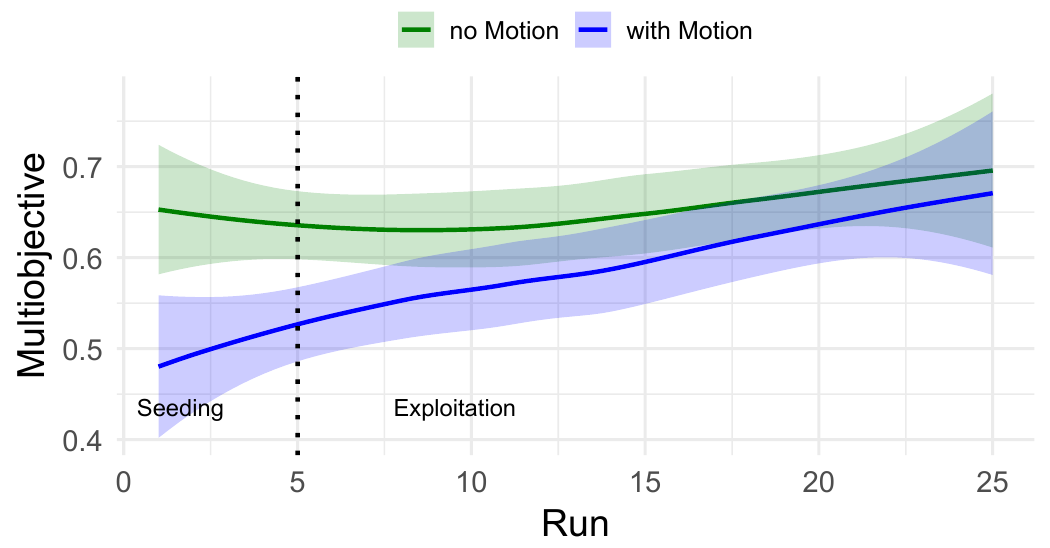}
    \caption{Optimization Process of the averaged normalized objectives during the 30 runs}\label{fig:optimization_process_for_combined_data}
    \Description{This figure shows the increase of both groups for the combined multiobjectives}
\end{figure}

In \autoref{app:optimization_process}, we illustrate the optimization process across all participants for each objective. The results clearly show that most individual objectives improved over the course of the runs. Additionally, the aggregated and normalized scores for the combined objectives steadily increased throughout the optimization process, indicating that the iterative approach effectively enhanced the overall design outcomes. \autoref{fig:optimization_process_for_combined_data} shows the optimization process for the combined objectives. This was done by normalizing the ratings across all six objectives to a range of $[-1, 1]$. For the mental demand, the ratings were inverted to align with the overall optimization direction (see \autoref{sec:objective_functions}).

To evaluate the differences between the two groups, we employed an independent Bayesian t-test. Unlike traditional null hypothesis significance testing (NHST), Bayesian analysis offers a clear advantage: it directly quantifies the evidence supporting one hypothesis over another. As noted by \citet{williams_using_2017}, ``\textit{a Bayes factor analysis makes it clear when a set of observed data is more consistent with the null hypothesis than the alternative}''~\cite[p. 1]{williams_using_2017}. This approach is particularly advantageous as Bayesian methods are inherently designed to quantify evidence directly. In contrast, NHSTs are more centered on decision-making processes, such as rejecting or not rejecting a null hypothesis, rather than providing a directional measure of evidence~\cite{morey_philosophy_2016}. In contrast, the Bayes factor directly indicates the strength of the evidence. 

According to \citet{lee2013bayesian}, a Bayes factor less than 1 provides evidence in favor of the null hypothesis, suggesting equality between the groups. Conversely, a Bayes factor greater than 1 suggests evidence for a difference between the groups. The further the Bayes factor is from 1, the stronger the evidence supporting the respective hypothesis. If the Bayes factor is close to 1, the data do not strongly support either hypothesis.


\subsubsection{Immersion}
The descriptive data revealed that immersion with motion cues was rated higher (\m{21.3}, \sd{4.94}) than the simulation with only a VR setup (\m{18.5}, \sd{5.76}), with possible scores ranging from 4 to 28. A Bayesian independent t-test was conducted to examine the effect of the motion chair on immersion. However, the analysis yielded a Bayes factor of BF = 0.998±0.01\%, suggesting that the data provide no evidence for equality or difference between both groups. This indicates that there is no evidence regarding the effect of motion fidelity on immersion.

\subsubsection{Design Parameters}
\begin{figure*}[ht!]
\centering
\small

    \begin{subfigure}[c]{0.46\linewidth}
        \includegraphics[width=\linewidth]{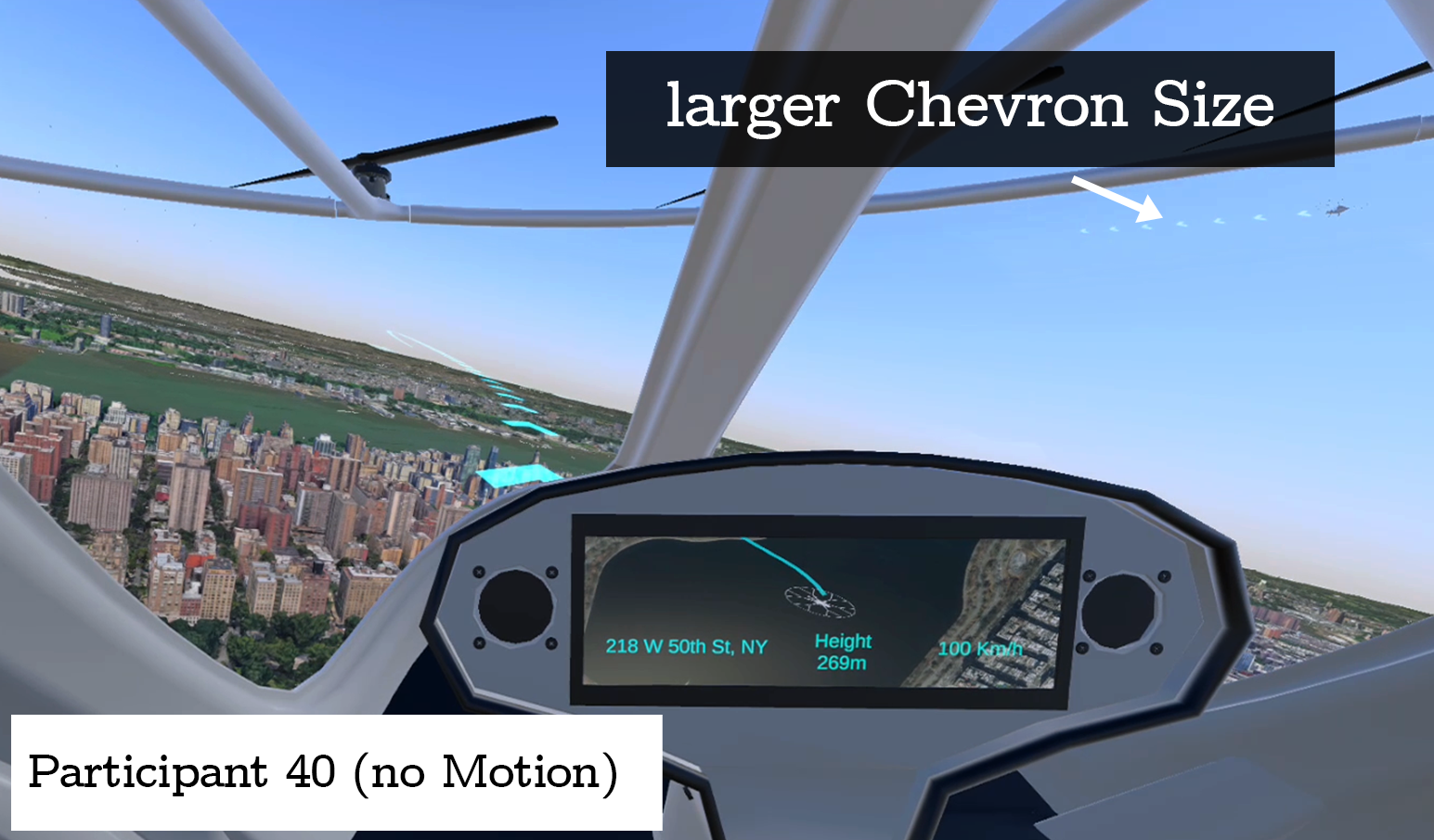}
        \caption{Pareto front visualization of participant 40 \textbf{without Motion}}\label{fig:optimized_design_no_motion}
        \Description{This figure depicts the optimized user interface for participant 40 (no Motion). The interface shows the ego flight trajectory, a map at the display, and additional information (altitude and speed). This UI does not include boundary boxes around other air taxis.}
    \end{subfigure}
    \hspace{1em}
    \begin{subfigure}[c]{0.46\linewidth}
        \includegraphics[width=\linewidth]{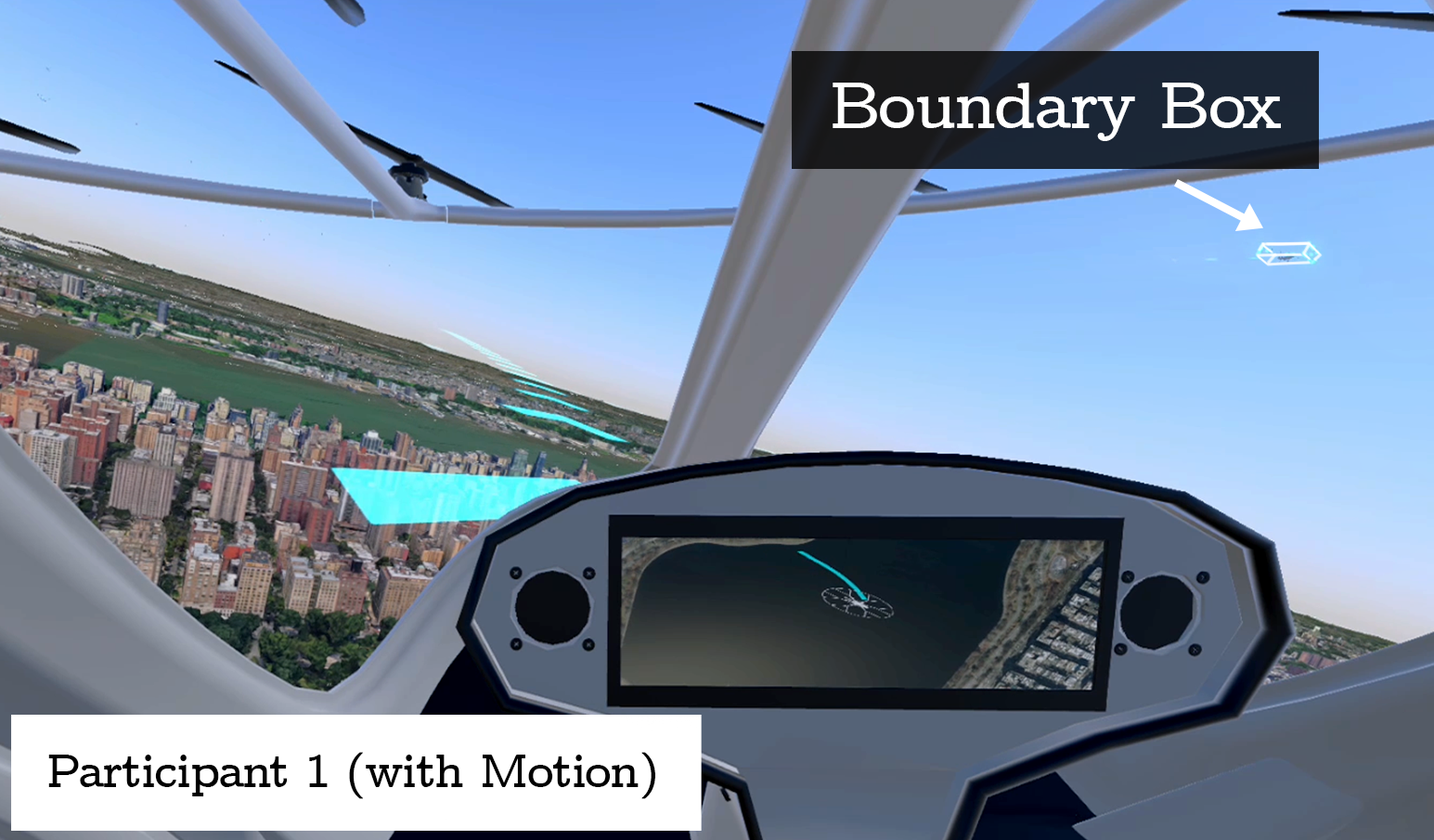}
        \caption{Pareto front visualization of participant 1 \textbf{with Motion}}\label{fig:optimized_design_no_motion}
        \Description{This figure shows the UI design for participant 1 (with Motion). It shows a similar flight trajectory and flight data as in 3a, but with smaller chevrons representing the other air taxis' trajectories. Additionally, boundary boxes are shown around other air taxis.}
    \end{subfigure}

   \caption{Exemplary optimized design parameters of Participant 40 (no Motion) and Participant 1 (with Motion). The concrete design parameters of these participants are plotted in \autoref{fig:design_parameters}. Strong to extreme evidence of differences was found for the boundary box and chevron size of other air taxis. All other differences of design parameters, such the additional information at the display are due to personal preferences but no evidence was found to support this difference between groups}~\label{fig:optimized_parameters}
   \Description{This figure shows the optimized design parameters for two examplary participants. Strong to extreme evidence was found for differences for the chevron size and the presence of boundary boxes around other air taxis}
\end{figure*}

\begin{figure*}[ht!]
    \centering
    \small
    \includegraphics[width=0.99\linewidth]{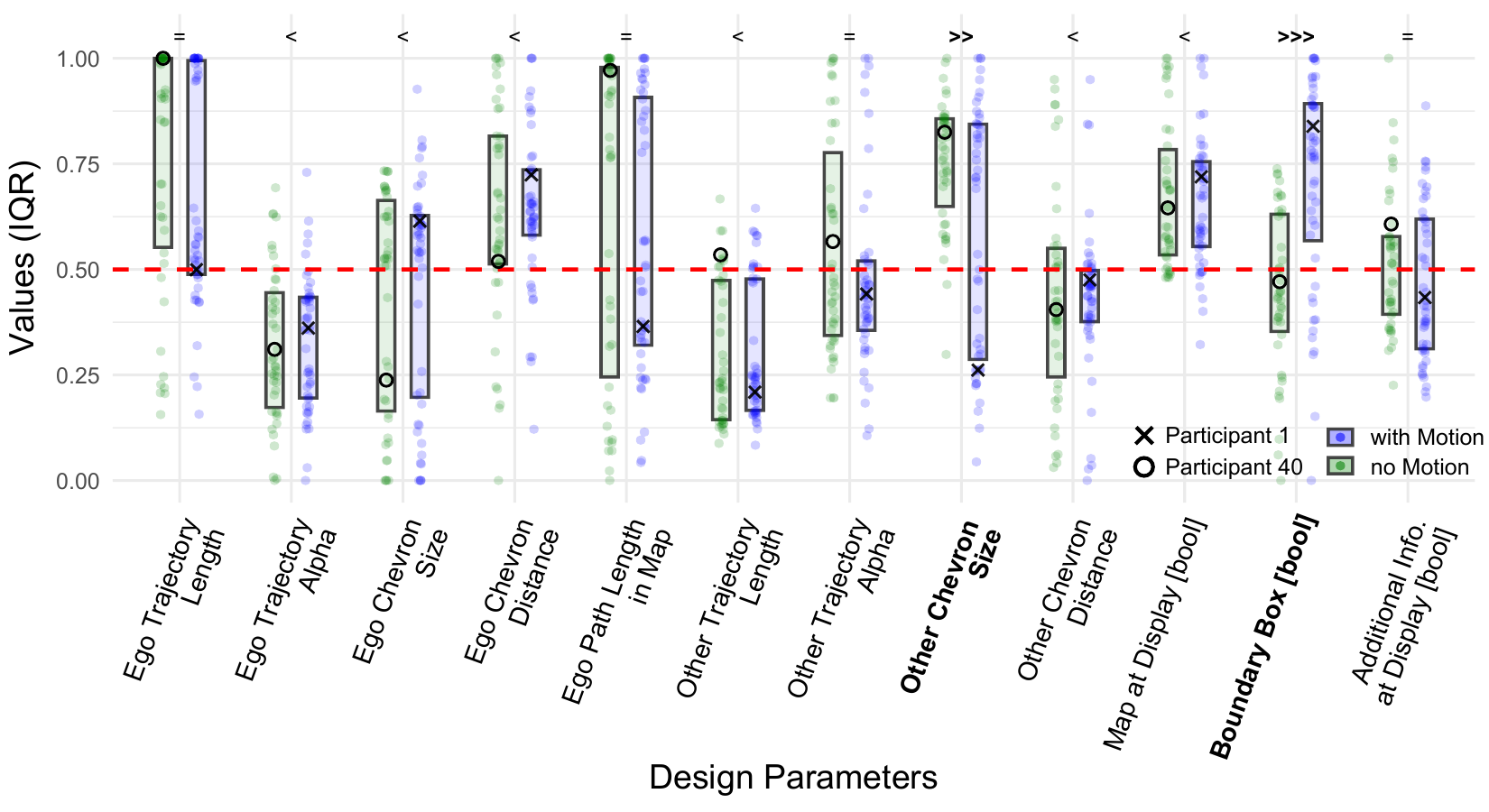}
    \caption{Comparison of the design parameters for both groups, using a Bayesian t-test. For both groups, the IQR is plotted. Further, the exemplary Pareto front designs of Participant 1 (with Motion) and Participant 40 (no Motion) are depicted (see \autoref{fig:optimized_parameters} for the visualization.
    The annotations are as follows defined by \citet{lee2013bayesian}: "$<<<$" for extreme evidence for equality (BF < 0.01), "$<<$" for strong or very strong evidence for equality (BF < 0.1), "$<$" for moderate evidence for equality (BF < 0.3), "$=$" for inconclusive or anecdotal evidence (BF between 0.3 and 3), "$>$" for moderate evidence for difference (BF > 3), "$>>$" for strong or very strong evidence for difference (BF > 10), and "$>>>$" for extreme evidence for difference (BF > 100). The dotted line indicates the threshold for the boolean design parameters to either be smaller or greater than 0.5}\label{fig:design_parameters}
    \Description{The plot shows that certain parameters, such as the size of chevrons and the inclusion of boundary boxes around other air taxis, have strong evidence for differences between the groups. In contrast, other parameters, like trajectory length or additional information display, show less or no difference or equalities, suggesting that motion cues had little impact on their optimization. The data for each design parameter is represented as IQR values, ranging from 0 to 1, indicating the preferred level of each parameter in both the motion and no-motion conditions.}
\end{figure*}

The results of our analysis for each design parameter are detailed in \autoref{tab:bf_parameters} and plotted in \autoref{fig:design_parameters}, which includes the IQRs for better representation of variability. While methods exist for visualizing multi-dimensional Pareto fronts (e.g., \cite{chiu_hyper-radial_2008, agrawal_intuitive_2004}), our study involves a unique Pareto front for each participant, making it challenging to combine these individual optimal designs into a single, unified UI. Hence, we present two representative Pareto designs—one from Participant 1 (motion condition) and another from Participant 40 (no-motion condition)—in \autoref{fig:optimized_parameters}. These examples were selected because most of their design parameter values mostly fall within the IQR of their respective groups (Participant 1: 11 out of 12 parameters; Participant 40: 9 out of 12 parameters are inside the IQR). These examples provide an illustrative comparison of one optimized design for each condition.

The parameter related to the chevron size of the other air taxis shows strong evidence for a difference between the two groups. This suggests that the optimized chevron size on the Pareto front for the "no Motion" group (\m{0.75}, \sd{0.15}) is larger than for the "with Motion" group (\m{0.59}, \sd{0.30}). There was also extreme evidence of a difference in the presence of a boundary box around the other air taxis (\m{0.45}, \sd{0.19} for the "no Motion" group and \m{0.71}, \sd{0.25} for the "with motion" group). This parameter is binary. Hence, as the value for "no Motion" is below the 0.5 threshold, this indicates that the optimal design for this group would not include the boundary box, while the optimal design for the "with Motion" group would include it, exceeding the threshold.

\begin{table*}[ht!]
\centering
\small
\caption{Results of Bayesian Analysis for Each Design Parameter including the Medians and IQRs}
\Description{This table shows the results for the Bayesian t-test comparing the no-motion group with the group with motion cues. It also presents the median and and IQRs for both groups}
\begin{tabular}{lllll}
\hline
\textbf{Design Parameter} & \textbf{BF (± \%)} & \textbf{no Motion Mdn (IQR)} & \textbf{with Motion Mdn (IQR)} & \textbf{Evidence} \\
\hline
Ego Trajectory Length & 0.41 ± 0.02\% & 0.75 (0.55, 1.00) & 0.74 (0.49, 0.99) & anecdotal equality \\
\textit{Ego Trajectory Alpha } & \textit{0.23 ± 0.02\%} & \textit{0.31 (0.17, 0.45)} & \textit{0.32 (0.20, 0.43)} & \textit{moderate equality} \\
\textit{Ego Chevron Size}      & \textit{0.26 ± 0.02\%} & \textit{0.41 (0.16, 0.66)} & \textit{0.41 (0.20, 0.63)} & \textit{moderate equality} \\
\textit{Ego Chevron Distance}  & \textit{0.22 ± 0.02\%} & \textit{0.61 (0.51, 0.82)} & \textit{0.66 (0.58, 0.74)} & \textit{moderate equality} \\
Ego Path Length in Map & 0.36 ± 0.02\% & 0.62 (0.25, 0.98) & 0.61 (0.32, 0.91) & anecdotal difference \\
\textit{Other Trajectory Length} & \textit{0.22 ± 0.02\%} & \textit{0.31 (0.14, 0.47)} & \textit{0.32 (0.17, 0.48)} & \textit{moderate equality} \\
Other Trajectory Alpha & 0.62 ± 0.02\% & 0.56 (0.34, 0.78) & 0.49 (0.36, 0.52) & anecdotal equality \\
\textbf{Other Chevron Size}    & \textbf{16.54 ± 0\%} & \textbf{0.76 (0.65, 0.86)} & \textbf{0.57 (0.29, 0.84)} & \textbf{strong difference} \\
\textit{Other Chevron Distance} & \textit{0.22 ± 0.02\%} & \textit{0.42 (0.25, 0.55)} & \textit{0.43 (0.38, 0.50)} & \textit{moderate equality} \\
\textit{Map at Display [bool]} & \textit{0.24 ± 0.02\%} & \textit{0.66 (0.53, 0.78)} & \textit{0.68 (0.55, 0.76)} & \textit{moderate equality} \\
\textbf{Boundary Box [bool]}   & \textbf{32473.69 ± 0\%} & \textbf{0.45 (0.35, 0.63)} & \textbf{0.71 (0.57, 0.89)} & \textbf{extreme difference} \\
Addn. Info. at Display [bool] & 0.51 ± 0.02\% & 0.49 (0.39, 0.58) & 0.47 (0.31, 0.62) & anecdotal equality \\
\hline
\end{tabular}
\label{tab:bf_parameters}
\end{table*}

\subsubsection{Questionnaire Ratings}\label{sec:questionair_ratings}
\begin{figure*}[ht!]
\centering
\small
    \begin{subfigure}[c]{0.16\linewidth}
        \includegraphics[width=\linewidth]{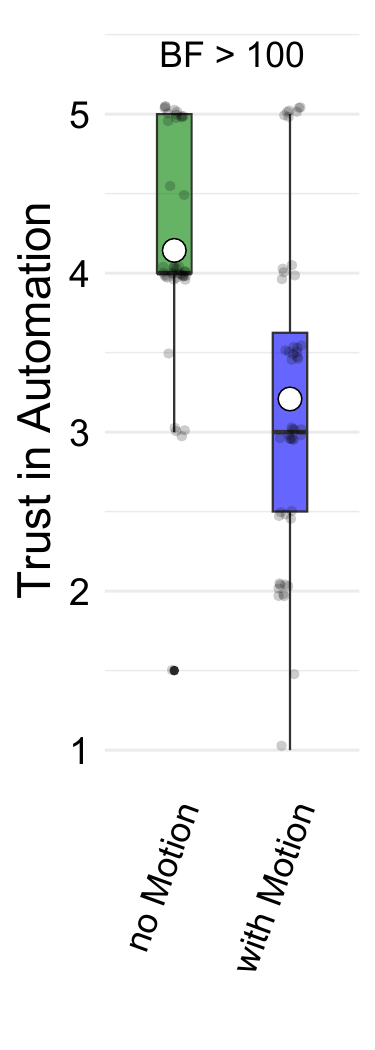}
        \caption{Trust}\label{fig:trust}
        \Description{Trust in Automation}
    \end{subfigure}
    \begin{subfigure}[c]{0.16\linewidth}
        \includegraphics[width=\linewidth]{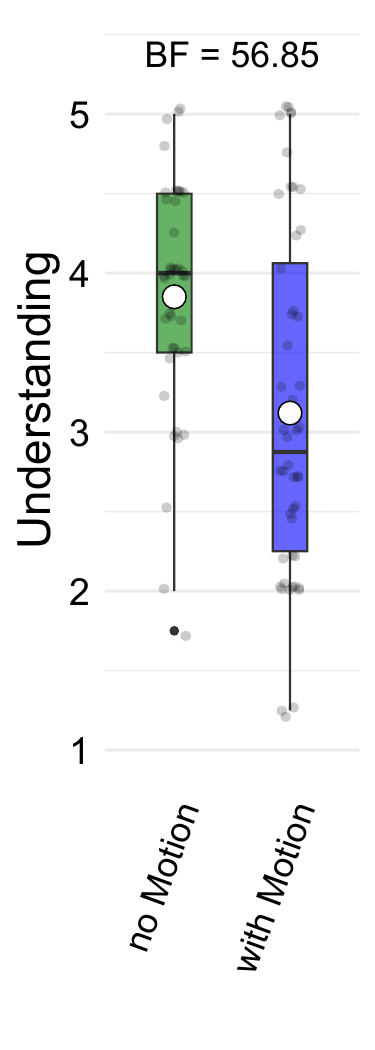}
        \caption{Understanding}\label{fig:understanding}
        \Description{Understanding}
    \end{subfigure}
    \begin{subfigure}[c]{0.16\linewidth}
        \includegraphics[width=\linewidth]{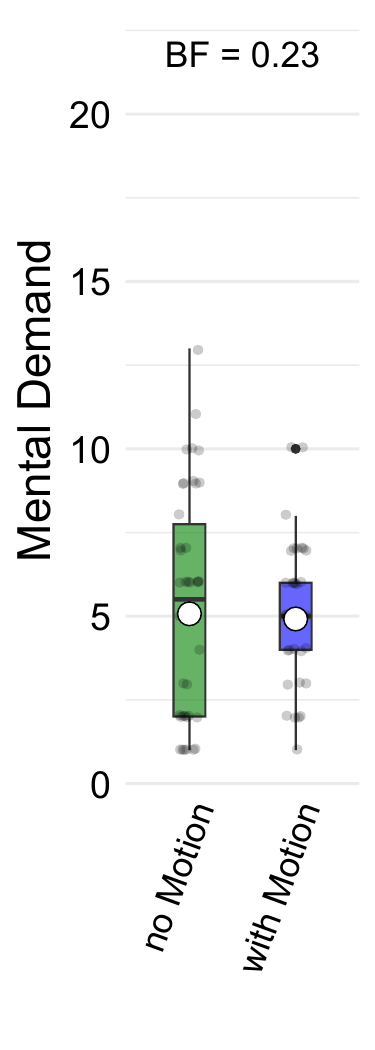}
        \caption{Mental Demand}\label{fig:mental_demand}
        \Description{Mental Demand}
    \end{subfigure}
    \begin{subfigure}[c]{0.16\linewidth}
        \includegraphics[width=\linewidth]{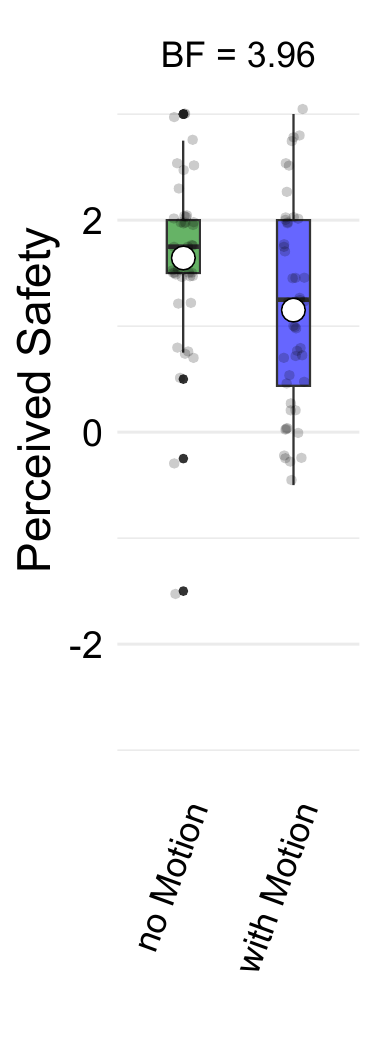}
        \caption{Perceived Safety}\label{fig:perceivedsafety}
        \Description{Perceived Safety}
    \end{subfigure}
    \begin{subfigure}[c]{0.16\linewidth}
        \includegraphics[width=\linewidth]{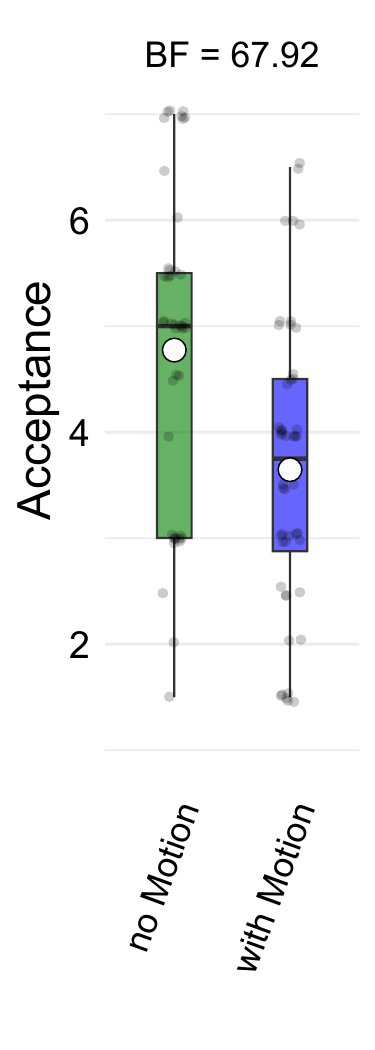}
        \caption{Acceptance}\label{fig:acceptance}
         \Description{Acceptance}
    \end{subfigure}
    \begin{subfigure}[c]{0.16\linewidth}
        \includegraphics[width=\linewidth]{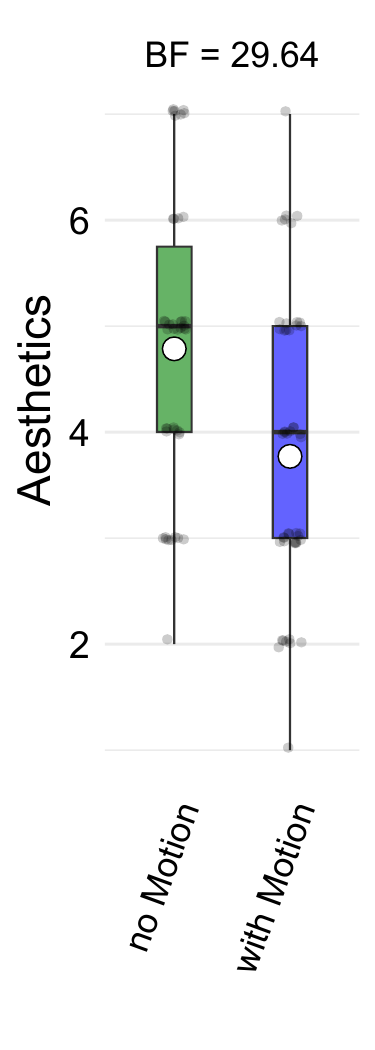}
        \caption{Aesthetics}\label{fig:Aesthetics}
        \Description{Aesthetics}
    \end{subfigure}

   \caption{Rating for the subjective questionnaires comparing both groups: no Motion and with Motion of all Pareto optimal values. The Bayes factor shows the trend towards equality (<1) and difference (>1)}~\label{fig:questionair_data}
   \Description{This figure shows the boxplots for comparing the questionnaire ratings (objectives) for both groups. The figure consists of six box plots (a-f), each representing a different aspect of user experience measured in the study: Trust in Automation, Understanding, Mental Demand, Perceived Safety, Acceptance, and Aesthetics.}
\end{figure*}

We analyzed the mean ratings from the questionnaires of all participants whose design parameters were on the Pareto front. 
We then compared these ratings between the group with motion cues and the group without motion cues. Refer to \autoref{fig:questionair_data} for a comparison overview.

\paragraph{Trust in Automation}
The Bayesian analysis of the trust measure resulted in a BF of 3970.14 ± 0\%, suggesting extreme evidence for a difference between the group without the motion cues (\m{4.14}, \sd{0.73}) and with motion cues (\m{3.21}, \sd{1.02}, see \autoref{fig:trust}).

\paragraph{Understanding}
For the understanding measure, the analysis yielded a BF of 56.85 ± 0\%, providing very strong evidence for the differences between the two groups. For the optimized design parameters on the Pareto front, the motion setup yielded a lower understanding (\m{3.12}, \sd{1.11}) than the setup without the motion chair (\m{3.85}, \sd{0.73}, see \autoref{fig:understanding}).

\paragraph{Mental Demand}
The mental demand measure showed a BF of 0.23 ± 0.02\%, indicating moderate evidence for equality between the groups. Hence, the design parameters in the Pareto front in the setup with motion yield similar ratings (\m{4.92}, \sd{2.01}) compared to the ones without motion (\m{5.07}, \sd{3.49}, see \autoref{fig:mental_demand}).

\paragraph{Perceived Safety}
The analysis of perceived safety resulted in a BF of 3.96 ± 0.01\%, with moderate evidence in favor of a difference between the groups. The Pareto front design was rated lower for the group with motion (\m{1.15}, \sd{0.95}) than the one without motion (\m{1.64}, \sd{0.84}). See \autoref{fig:perceivedsafety} for more details.

\paragraph{Acceptance}
The acceptance had a BF of 67.92 ± 0\%, suggesting very strong evidence in favor of a difference between the group without motion (\m{4.77}, \sd{1.53}) and with motion (\m{3.65}, \sd{1.36}, see \autoref{fig:acceptance}).

\paragraph{Aesthetics}
The Bayesian analysis yielded a BF of 29.63 ± 0\% for the aesthetics measure, indicating moderate evidence favoring the group's difference. The optimal Pareto front design parameters yield a lower rating for the group with motion  (\m{3.77}, \sd{1.43}) compared to the group without motion (\m{4.79}, \sd{1.39}, see \autoref{fig:Aesthetics}).

\subsection{Correlation between the Objectives}
\begin{figure}[ht!]
    \centering
    \small
    \includegraphics[width=0.97\linewidth]{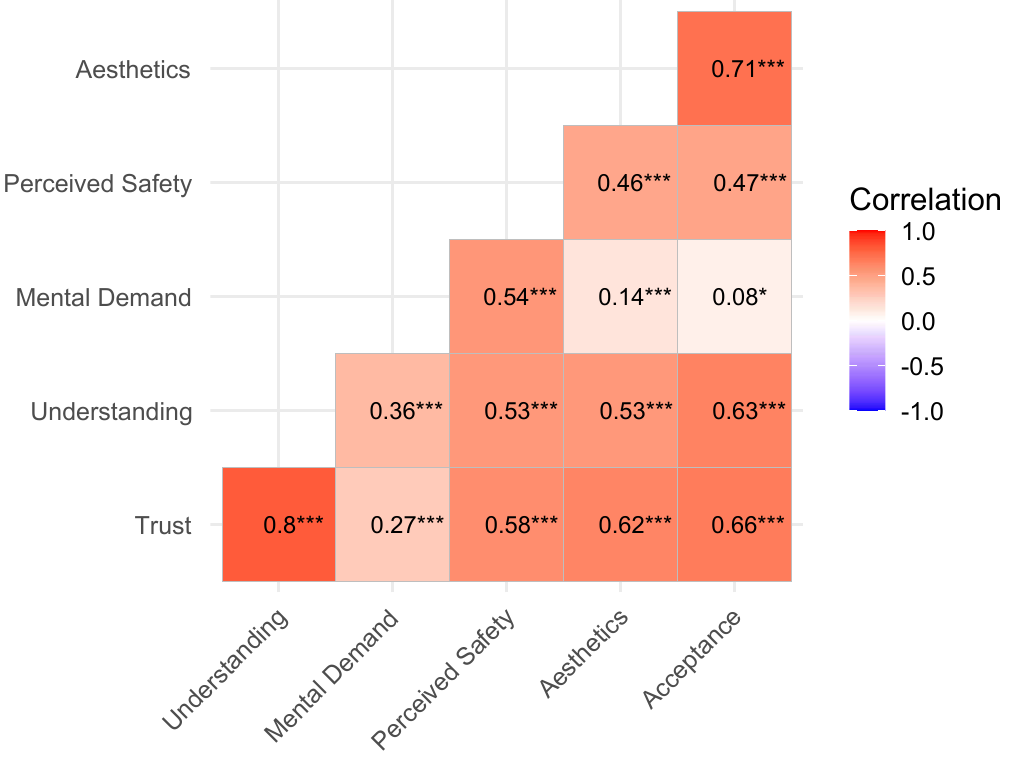}
    \caption{Correlation of all Objectives shows that most of them correlate and are not trade-offs to each other}\label{fig:correlation}
    \Description{This figure is a correlation heatmap, displaying the relationships between the objectives used in the study: Trust, Understanding, Mental Demand, Perceived Safety, Aesthetics, and Acceptance. The figure uses a color gradient, with lighter shades representing weaker correlations and darker shades indicating stronger correlations. The correlation values range from -1.0 to 1.0, where positive values indicate a positive correlation and negative values indicate a negative correlation.
    Strong positive correlations are observed between Trust and Understanding (0.80), and between Aesthetics and Acceptance (0.71). Weak correlations are observed between Mental Demand and Aesthetics (0.14) and Acceptance (0.08).}
\end{figure}

MOBO identifies optimal design parameters that balance multiple objectives along the Pareto front, ensuring that no single objective can be improved without compromising others~\cite{marler_survey_2004}. To assess whether there are trade-offs between these objectives, we calculated the correlations among all objectives (see \autoref{fig:correlation}). This analysis helps us understand how changes in one objective might influence others.
The results show that all correlations were statistically significant. In particular, trust and understanding had a strong positive correlation ($r=0.8$), as did acceptance and aesthetics ($r=0.71$), meaning improvements in one were often associated with improvements in the other. In contrast, the correlations between Mental Demand and Aesthetics ($r=0.14$) and Mental Demand and Acceptance ($r=0.08$) were weak. The remaining objective pairs demonstrated moderate positive correlations.

\section{Discussion}
Automated UAM is expected to reduce urban congestion, with air taxis expected to operate in major cities within the next decade~\cite{EuropeanUnionAviationSafetyAgency.2021, VolocopterGmbH_WhitePaper}. However, successfully adopting UAM will depend on passenger trust and perceived safety in this novel mode of transport~\cite{AlHaddad.2020}. Research has shown that effective UI designs can significantly increase passengers' trust and perceived safety~\cite{colley2023comefly, valente_towards_2024}. However, to develop and research optimal UI designs for air taxis, it is crucial to use motion-fidelity simulators that accurately reflect the motion conditions of the UAM~\cite{Edwards2020, meinhardt2023up}. While some studies have investigated the motion fidelity of simulators in the field of AVs~\cite{yeo_toward_2020, hock_introducing_2022, colleySwiVRCarSeatExploringVehicle}, there remains limited understanding of the effects of motion fidelity on passengers and the UI in the field of UAM. Therefore, this research was driven by two RQs:

\smallskip
\smallskip

\noindent\fcolorbox{black}{black!10}{\textbf{RQ1}} \textit{What are the characteristics of an optimized UI design for automated air taxis that enhance passengers' trust, perceived safety, mental demand, understanding, acceptance, and aesthetics?}

\smallskip
\smallskip

\noindent\fcolorbox{black}{black!10}{\textbf{RQ2}}\textit{ How does the motion fidelity of the simulation affect the design parameters and user's effects for a UI in automated UAM?}

\smallskip
\smallskip

To explore these RQs, we conducted a between-subject study with N=40 participants. One group experienced a simulated air taxi flight using a 3-DoF motion chair combined with VR (n=20), while the other group experienced the same simulation in VR without motion cues (n=20). For both groups, we applied MOBO to iteratively optimize the UI designs based on participants' ratings of trust, perceived safety, understanding, mental demand, acceptance, and aesthetics~\cite{chan2022bo}. This approach allowed us to compare user ratings and the resulting optimal UI designs for both motion and no-motion conditions on the Pareto front, where improving one objective inherently requires compromising another~\cite{marler_survey_2004}.

\subsection{Motion Fidelity Decreases Trust, Understanding, and Acceptance}\label{sec:discussion_motion_fidality}
Our analysis found no evidence of a difference or equality in immersion between participants who experienced motion cues and those who did not. In contrast, \citet{yeo_toward_2020} found that motion platforms in VR significantly increased presence. While immersion refers to the technical quality of the virtual environment, presence describes the psychological feeling of being in a (virtual) place~\cite{slater_framework_1997}. 
The lack of difference in immersion in our study may be attributed to the fact that none of the participants had previously experienced a real flight in an automated air taxi, making it difficult for them to rate the immersion of the simulations accurately. This raises the question of whether valid studies can be conducted in UAM contexts without participants having firsthand experience with air taxi movements. In contexts like AVs, most users already have some experience with driving or being passengers in vehicles, which makes it easier to simulate and measure realistic user responses. However, in UAM, there is no widespread base of experience, which can weaken the reliability of participant responses in simulations. 

We, however, found strong to extreme evidence that motion fidelity negatively impacted trust, understanding, and acceptance (see \autoref{sec:questionair_ratings}). This aligns with concerns about automated air taxis in prior (market) studies~\cite{EuropeanUnionAviationSafetyAgency.2021, AlHaddad.2020, Edwards2020}, suggesting that participants' lower ratings in the motion group may more accurately reflect how passengers will respond to future real-world automated air taxi scenarios. While the MOBO process successfully optimized UI design to address these effects to some extent, resulting in slightly adjusted UI designs, it could not fully compensate for the negative impact of motion on trust, understanding, and acceptance. Although the MOBO process optimized UI design to mitigate these effects (see \autoref{app:optimization_process}), the optimized interface was unable to reach the level of trust, understanding, and acceptance with the no motion condition. This aligns with the theory of trust by \citet{lee_trust_2004}, who state that trust in automation depends on multiple factors beyond UI design, named environmental context including the user's self-confidence. Similarly, \citet{hoff2015trust} proposed that trust also depends on the pre-knowledge of the users. This suggests that while interface improvements can address some aspects of user experience, they alone cannot resolve fundamental trust deficits caused by the motion simulation.

Nevertheless, when studying user effects in the UAM context, we recommend using motion fidelity simulators to closely replicate participants' real-world ratings when actual/absolute ratings are necessary. However, since most air taxi interface design studies focus on comparing different interface designs~\cite{colley2023comefly, valente_towards_2024}, they primarily require relative comparisons rather than absolute real-world ratings. For these comparative studies, simulating exact real-world user ratings during an air taxi flight is less critical, and even a monitor study such as by \citet{valente_towards_2024} might be sufficient.

\subsection{Optimized Design Choices in Automated Air Taxi Interfaces}
The comparison of optimized UI design parameters on the Pareto front revealed only minor evidence for differences or similarities between the motion and no-motion conditions. However, two strong/extreme evidences for differences emerged. First, there was strong evidence that the size of the chevrons representing other air taxis was smaller with motion than without. Second, there was extreme evidence suggesting that, in the motion condition, participants preferred to visualize boundary boxes around other air taxis, while in the no-motion condition, participants' ratings suggested not to display these boxes. Although these differences were subtle, as the IQR of the no-motion group overlaps with the 0.5 threshold line (see \autoref{fig:design_parameters}), the preference for boundary boxes in the motion condition aligns with participants' reduced trust in this scenario. Participants experiencing motion likely felt a stronger need for visual confirmation that other air taxis were detected, reflecting the compensatory role of interface elements in addressing reduced trust in motion fidelity contexts. Further, participants' preferences for smaller chevrons in the motion group suggest that larger chevrons of other air taxis may be distracting when motion is added. However, given that simulations aim to replicate real-world conditions as closely as possible—-and in reality, motion would always be present-—we argue that boundary boxes should be included in all simulations, regardless of whether motion fidelity is present. Similarly, chevron sizes for other air taxis should remain small in all conditions to avoid distraction.

Beyond these specific differences, the overall impact of motion fidelity on UI design was minimal, suggesting that motion has only limited influence on most design parameters. Also, visually comparing individuals' UI designs (see \autoref{fig:optimized_parameters}) yielded only minimal perceivable differences. We, however, observed that the starting point for all objectives was already relatively high (see \autoref{app:optimization_process}). Hence, It seems that participants found any UI combination to be sufficient as long as some form of visualization was present. This is further supported by the fact that none of the design parameters approached a value of zero, indicating that all visual elements were considered necessary to some extent.
This aligns with findings from Colley and Meinhardt et al.~\cite{colley2023comefly}, where even simple displays significantly improved users' trust ratings compared to having no visualization at all.

\subsection{User Interface Optimization Process}
The relatively high variation in the UI design parameters on the Pareto front indicates that a single, universally optimized design was unsuitable for the participants. While some design parameters, such as the alpha of the ego trajectory and the length of the other air taxis' trajectories, showed relatively low variance and consistent preferences among participants, other values showed high variation. 
This suggests that a one-size-fits-all approach to UI design may not be effective for automated air taxis. Even though most of the objectives constantly increased during the optimization process for both groups, creating personalized UI designs tailored to the individual user's needs may be more beneficial than focusing on a single optimized design. This is consistent with instrumental personalization, as described in the framework by \citet{fan_what_2006} on personalization, which emphasizes adapting tools and systems to users' individual needs to improve efficiency and effectiveness, making it a relevant concept for UAM interface design. This approach has already been explored in the automotive domain, where \citet{normark2015design} found that \textit{``[...] personalizable vehicle interfaces can improve both the usefulness and the user experience with the product''}~\cite[p. 744]{normark2015design}. One could extend this concept by considering personalized flight styles, as previous research for AVs has already shown that personalizing driving styles (such as driving speed and acceleration) could increase passengers' trust and comfort~\cite{sun_exploring_2020}. Moreover, this preference for personalization extends beyond AVs. Emergency helicopter pilots have also preferred personalized UIs during their flights~\cite{janetzko_what_2024}. This suggests that such a tailored approach could be equally beneficial for inexperienced passengers in automated air taxis. However, design decisions must also consider practical constraints beyond trust and acceptance, such as cost, legal restrictions, or weight limitations of the hardware. Since UI design has a limited effect on trust compared to other influencing factors~\cite{lee_trust_2004} (see \autoref{sec:discussion_motion_fidality}), a degree of creative freedom for designers should remain, allowing them to balance functional and aesthetic priorities. This freedom allows them to balance functional needs with aesthetic and practical considerations. One way to integrate this creative flexibility with BO is by restricting the algorithm's design space, enabling it to optimize within predefined boundaries that align with broader design priorities. For instance, \citet{mo_cooperative_2024} suggest using forbidden regions to exclude undesirable search-space areas before and during optimization, ensuring that results remain aligned with user needs and designers' goals.

Moreover, the feasibility of personalization depends on the context. While personalized interfaces may work well for individual air taxis, future visions of UAM also include shared Air Metros, described as \textit{``[...] similar to today's public transportation systems such as subways and buses [...]''}~\cite[p. 4]{pak_can_2024}. In such scenarios, where multiple passengers share the same space, tailoring the interface to a single user is impractical. Instead, an standardized interface design that simultaneously accommodates the needs of multiple passengers would be necessary, ensuring usability and functionality for all.

\subsection{Trade-off in the Objectives}\label{sec:trade-off}
MOBO aims to find optimal design parameters that balance multiple objectives along the Pareto front, ensuring that improving one objective does not excessively compromise another~\cite{marler_survey_2004}. This is particularly useful when trade-offs between objectives exist, such as spatial error versus completion time in 3D touch interaction~\cite{chan2022bo}. However, our analysis revealed significant correlations among all six objectives, indicating that they are not in conflict with each other. This finding suggests that the set of objectives may be redundant in this study. Although selected based on prior UAM studies~\cite{colley2023comefly, meinhardt_wind_2024, valente_towards_2024}, future research could streamline them by focusing on the most distinct ones. Since trust and understanding, as well as acceptance and aesthetics, are strongly correlated (see \autoref{fig:correlation}), choosing one from each pair would simplify future studies. Specifically, we recommend prioritizing acceptance over aesthetics, as acceptance relies on a validated scale~\cite{van1997simple}, whereas the aesthetics metric, as introduced in~\cite{colley2023comefly}, was self-created without undergoing validation. Similarly, we propose focusing on trust directly rather than understanding, as understanding is only an underlying dimension of trust in automation~\cite{korber2018theoretical}.

Mental demand, with weak correlations to other objectives, should be retained as it seems to capture a distinct user effect. Lastly, perceived safety could be treated as a supplementary objective, as it is only moderately correlated to trust, mental demand, or acceptance. While these are preliminary observations, future research should further explore and validate the potential reduction and refinement of objectives in UAM studies to ensure more efficient design optimization.
For example, relevant psychological models, such as \textit{trust in automation} by \citet{korber2018theoretical}, suggest that understanding is an underlying dimension of trust, and therefore, they assess these dimensions separately in their questionnaire. While this distinction is important for in-depth psychological analysis, we argue that in the context of the MOBO process, due to the strong correlation between trust and understanding, separating them may not be more effective for optimization purposes.

\subsection{Practical Guidelines for Interface Designs in Automated Urban Air Mobility}
The findings from our study offer several practical insights for future research regarding future UIs for automated air taxis. Below, we provide key guidelines:

\begin{itemize}[leftmargin=*, itemsep=0pt]
    \item \textbf{Use Motion Cues for Realism, But VR Alone Suffices for UI Comparisons:} When designing UAM interfaces, including motion fidelity simulations (e.g., using a motion chair) is crucial when requiring closer-to real-world user ratings. However, since motion fidelity did not have a strong impact on the UI design, it seems that only VR or even monitors without motion cues is sufficient for studies that compare UI designs.

    \item \textbf{Focus on Personalized Interfaces for Automated Air Taxis:} Due to the high variability in user preferences for the optimized UI, a one-size-fits-all approach for air taxi interfaces may not be effective. Instead, focus on developing personalized interface solutions that allow users to personalize elements based on their individual information needs, such as visualization of the ego trajectories.
  
    \item \textbf{Prioritize Providing Visual Information Over Fine-Tuning Details:} The high starting point for most objectives during the optimization process, especially in the seeding phase, suggests that any form of visualization in the interface is beneficial to contribute to passenger's trust, understanding and acceptance. Hence, rather than focusing on minor design adjustments, such as the exact length of the ego trajectory, ensuring the interface provides sufficient information to foster trust is more effective.

\end{itemize}

\subsection{Limitations and Future Work}
While we compared the effects on user's effect and optimal UI design for 3-DoF motion and without motion, we did not include more simulators into this study as done by \citet{yeo_toward_2020}. Hence, while we can derive our findings towards motion fidelity due to a 3-DoF motion chair, we cannot tell if other simulators might have yielded other results.
Further, we did not assess simulator sickness. Yet, previous work has shown that adding motion fidelity to simulations can reduce simulator sickness by better aligning physical and visual perceptions of motion in flight simulators~\cite{kennedy_simulator_1993, chen_effect_2020}. Conversely, mismatches between physical motion and visual cues can increase simulator sickness~\cite{mccauley_effects_1990}. In our study, the movements of the 3-DoF motion chair were precisely matched to the air taxi’s simulated motions, which likely minimized simulator sickness in the motion condition compared to the no-motion condition. Although simulator sickness was not the main focus of this study, evaluating it in future research could help clarify whether variations in UI evaluations are influenced by the presence of motion cues.

Additionally, we focused on subjective ratings as the primary objective for optimizing the UI design parameters. However, similar to the approach by \citet{chan2022bo}, incorporating objective measures such as real-time physiological data (e.g., heart rate or galvanic skin response) could offer a more holistic view of user experience. This integration would allow future studies to create more comprehensive optimization criteria for air taxi interfaces by combining both subjective feedback and objective physiological responses. Additionally, factors beyond user-centered metrics should be considered when optimizing UI designs for air taxis. For instance, hardware weight limitations or compliance with legal regulations may also influence design decisions. These practical considerations might conflict with user-preferred designs, highlighting the need for balanced trade-offs in future research. Nonetheless, the subjective ratings in this study were answered on validated questionnaires (trust~\cite{korber2018theoretical}, understanding~\cite{korber2018theoretical}, perceived safety~\cite{faas2020longitudinal}, mental demand~\cite{hart1988development}) or inspired by related work (acceptance~\cite{van1997simple}, aesthetics~\cite{colley2023comefly}). While valid,  future work should explore the relationship between these dimensions and more targeted subaspects of trust. It is also important to consider the potential redundancy of some questionnaire items due to strong correlations between certain measures (see \autoref{sec:trade-off}).

\section{Conclusion}
In this study, we explored the impact of motion fidelity in VR-simulated air taxi flights on passengers' effects and their interface design preferences using the MOBO approach. We conducted a between-subjects study with N=40 participants, divided into two groups of n=20, where one group experienced the simulation with motion cues using a 3-DoF motion chair, while the other group experienced it without motion cues, just in VR. 

Our findings reveal that while motion fidelity had minimal influence on most UI design parameters, there was evidence that it reduced users' trust, understanding, and acceptance, highlighting its importance in future UAM studies. However, the lack of differences in immersion may be attributed to participants' lack of experience with real air taxi flights, which makes it difficult for them to rate the immersion of the simulations accurately. 
Moreover, the relatively high variance in the optimal design preferences on the Pareto front suggests that a one-size-fits-all approach may not be suitable for UI design in automated UAM contexts. Hence, although we were able to show that most of the objectives constantly increased during the optimization process, personalized user interfaces tailored to individual preferences may provide better user experiences. 
Finally, due to the strong correlation between objectives like trust/understanding and acceptance/aesthetics, future studies may simplify the evaluation by focusing on fewer metrics. Optimizing one often improves others, making it unnecessary to evaluate all independently in the MOBO process. 

\begin{acks}
We would like to thank Johannes Schöning for his valuable mental support during the \textit{CHITogether 2024~\cite{scott2024doing}} in St. Gallen
\end{acks}

\bibliographystyle{ACM-Reference-Format}
\bibliography{sample-base}

\appendix
\onecolumn

\section{Optimization Process}\label{app:optimization_process}

\begin{figure*}[th]
\centering

    \begin{subfigure}[c]{0.45\linewidth}
    \includegraphics[width=\linewidth]{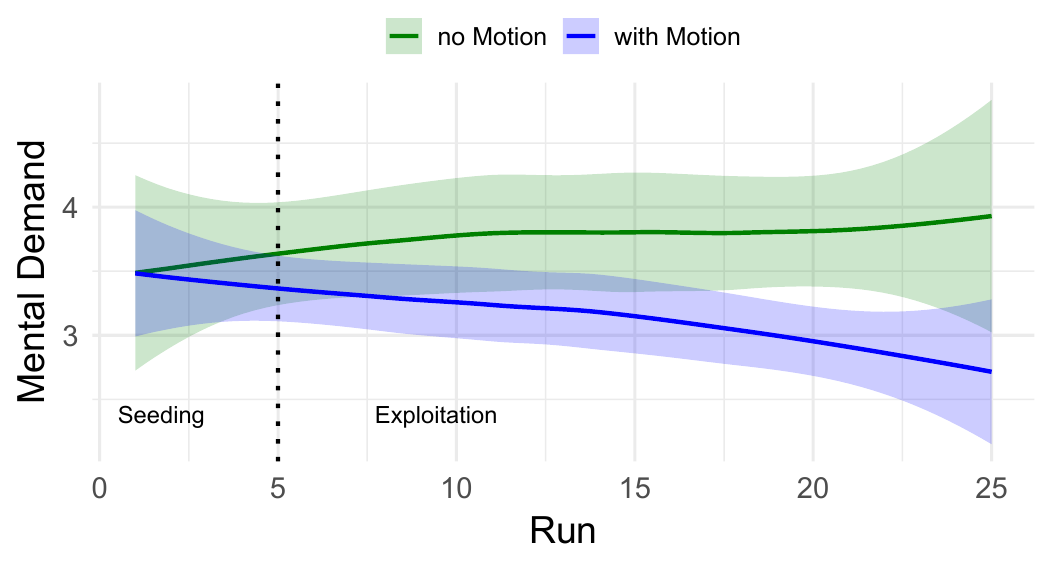}
        \caption{Mental demand}
        \Description{This figure shows the optimization process for mental demand. One can see that the mental demand for the no-motion group stays the same or even increases while the mental demand for the motion group decreases}
    \end{subfigure}
    \hspace{1cm}
    \begin{subfigure}[c]{0.45\linewidth}
    \includegraphics[width=\linewidth]{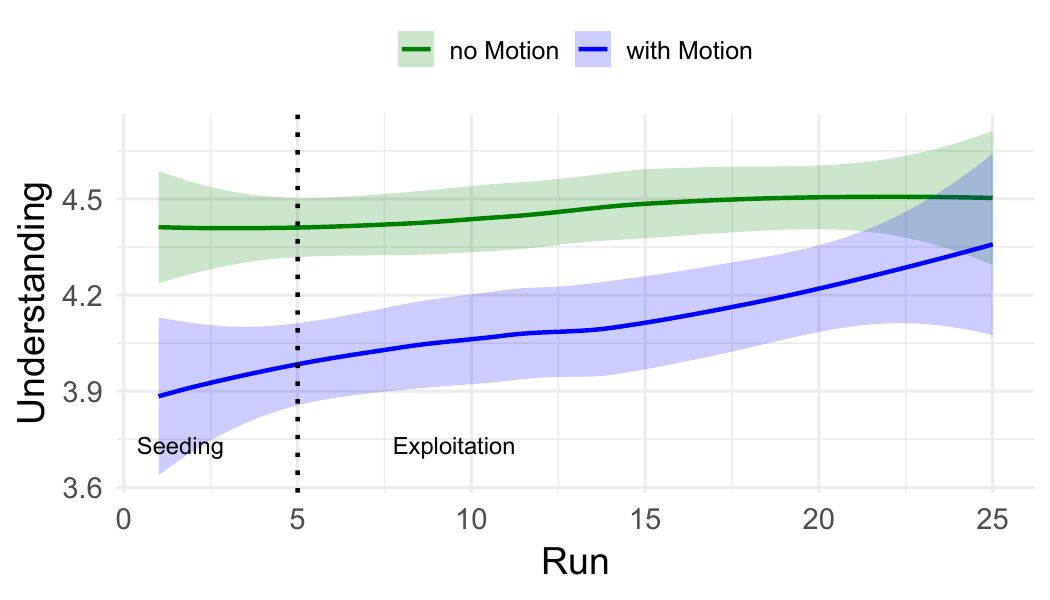}
        \caption{Understanding}
        \Description{This figure shows the optimization process for understanding. One can see that the understanding for the no-motion group stays the same  while the understanding for the motion group increases}
    \end{subfigure}
    \hspace{1cm}
    \begin{subfigure}[c]{0.45\linewidth}
    \includegraphics[width=\linewidth]{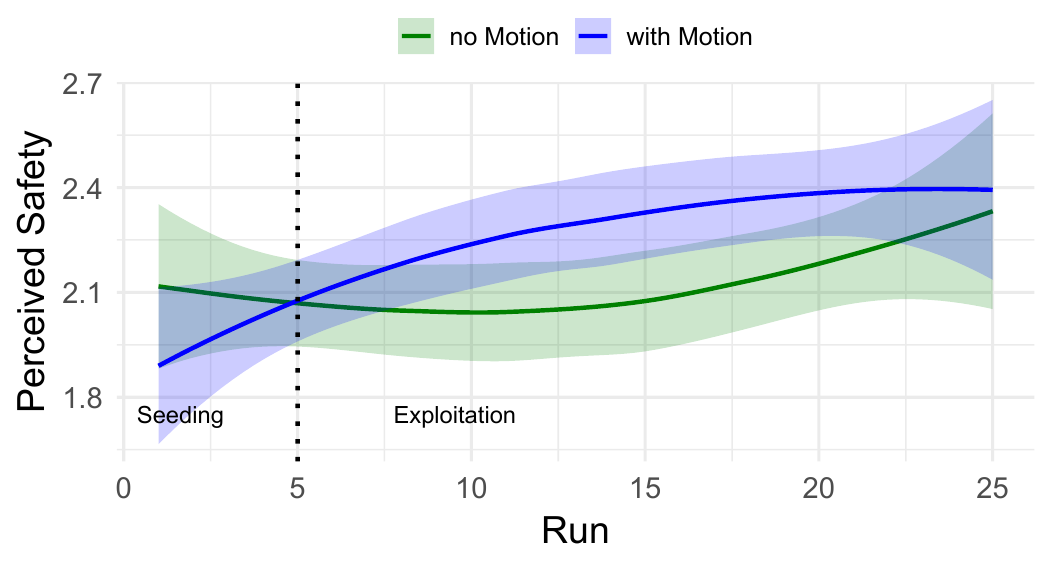}
        \caption{Perceived Safety}
        \Description{This figure shows the optimization process for perceived safety. One can see that the perceived safety for both groups increases}
    \end{subfigure}
    \hspace{1cm}
    \begin{subfigure}[c]{0.45\linewidth}
        \includegraphics[width=\linewidth]{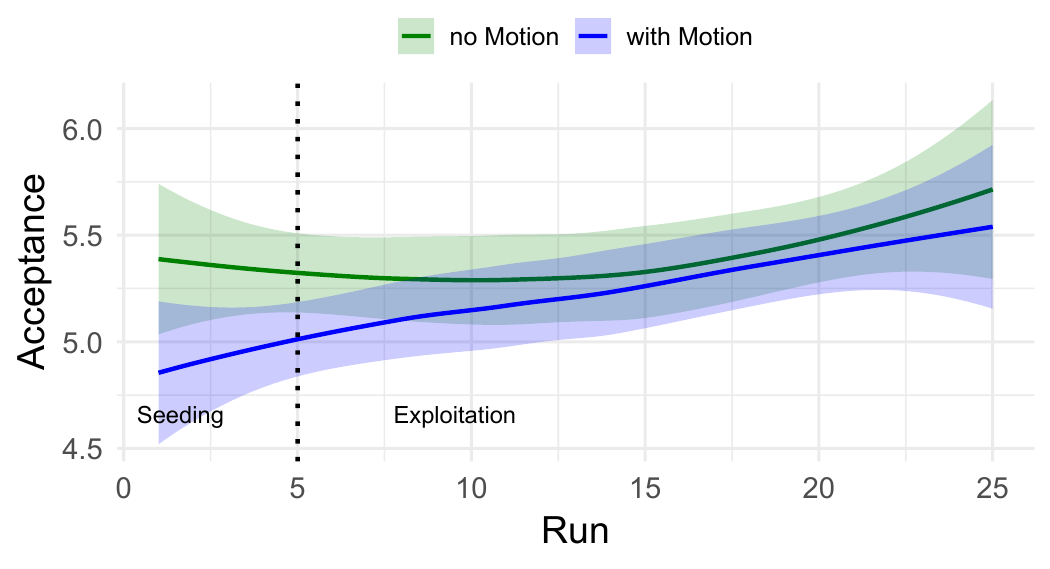}
        \caption{Acceptance}
        \Description{This figure shows the optimization process for acceptance. One can see that the acceptance for both groups increases in the exploitation phase}
    \end{subfigure}
    \hspace{1cm}
    \begin{subfigure}[c]{0.45\linewidth}
        \includegraphics[width=\linewidth]{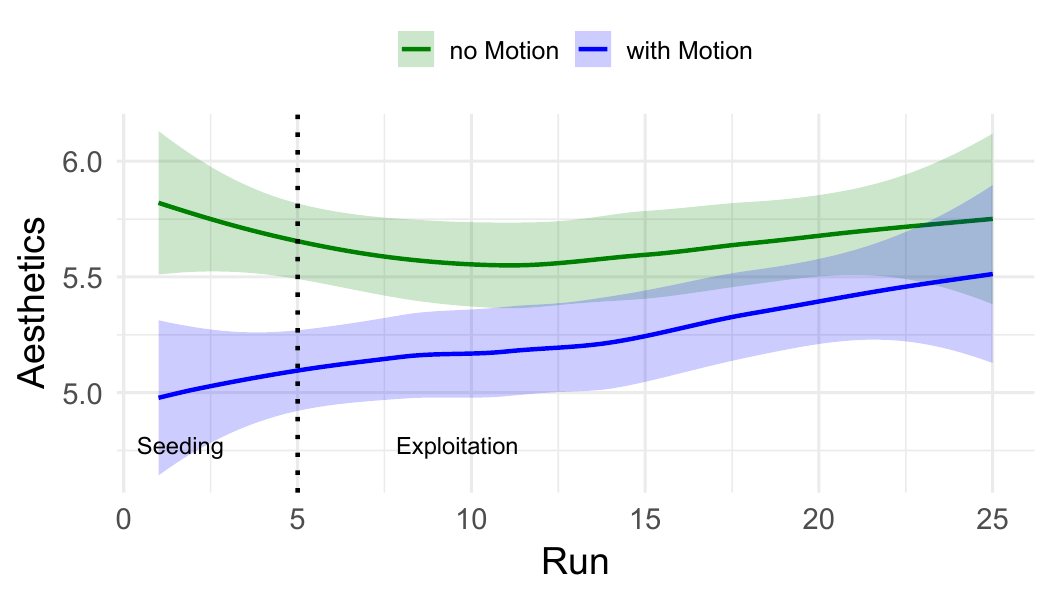}
        \caption{Aesthetics}
        \Description{This figure shows the optimization process for aesthetics. One can see that the aesthetics for both groups increase}
    \end{subfigure}
    \hspace{1cm}
    \begin{subfigure}[c]{0.45\linewidth}
        \includegraphics[width=\linewidth]{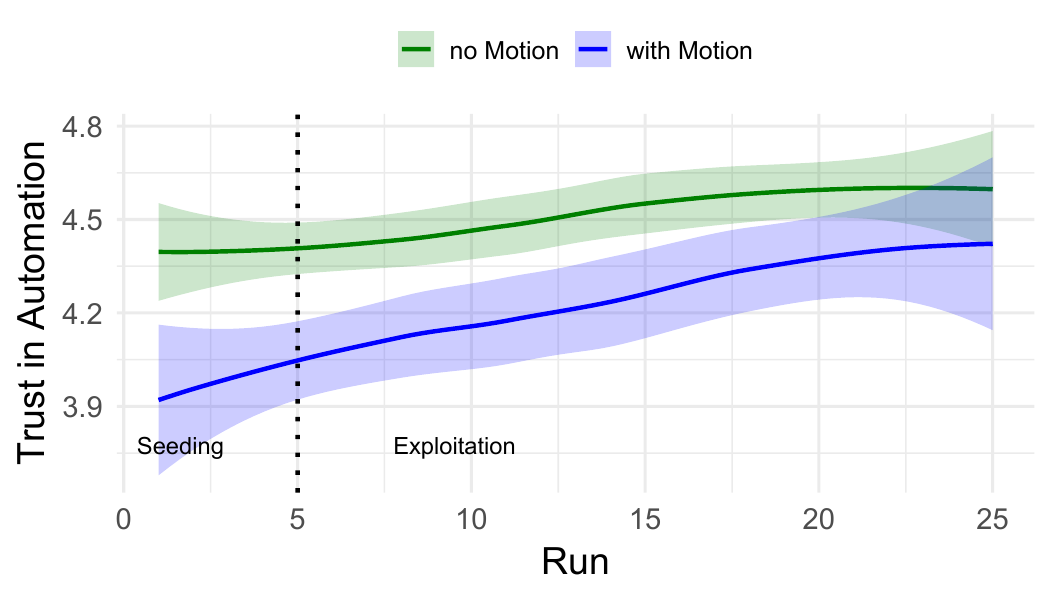}
        \caption{Trust}
        \Description{}
    \end{subfigure}
   \caption{Optimization process of each objective during the 30 runs}
   \Description{This figure shows the optimization process for each of the individual objectives during the 30 optimization runs.}
   \label{fig:optimization_process_all_questionairs}
\end{figure*}

\autoref{fig:optimization_process_all_questionairs} illustrates the optimization process across the individual objectives during the five seeding runs and 25 exploitation runs. Instead of optimizing the parameters to an individual optimum, we used MOBO to converge towards a Pareto front. This approach ensures that the UI design achieves a balanced optimization where no single objective can be improved without compromising another. 
We employed the Locally Estimated Scatterplot Smoothing (LOESS) method for data smoothing in the plots. Further, we set the span parameter to 1, which means that each point on the smoothed curve is influenced by the nearest 100\% of the data points. This choice ensures that the curve accurately represents the overall trend, effectively minimizing the impact of short-term fluctuations and noise.

\end{document}